\documentclass[preprint,3p,number,nopreprintline]{elsarticle}
\usepackage[utf8]{inputenc}
\usepackage[T1]{fontenc}
\usepackage{amsmath,amssymb,amsthm}
\usepackage{graphicx}
\usepackage{booktabs}
\usepackage{hyperref}
\usepackage{algorithm}
\usepackage{algpseudocode}
\usepackage{xcolor}
\usepackage{float}

\biboptions{sort&compress}
\journal{}
\hypersetup{
  hidelinks,
  hypertexnames=false,
  pdftitle={Orthogonal reparametrization of the Nelson--Siegel--Svensson interest rate curve model: conditioning, diagnostics, and identifiability},
  pdfauthor={Robert Flassig, Emrah G\"ulay, Daniel Guterding}
}
\newtheorem{proposition}{Proposition}[section]
\newtheorem{corollary}{Corollary}[section]
\newtheorem{remark}{Remark}[section]

\begin{document}
\begin{frontmatter}

\title{Orthogonal reparametrization of the Nelson--Siegel--Svensson interest rate curve model: conditioning, diagnostics, and identifiability}

\author[inst1]{Robert Flassig}
\ead{robert.flassig@th-brandenburg.de}
\author[inst2]{Emrah G\"ulay}
\ead{emrah.gulay@deu.edu.tr}
\author[inst1]{Daniel Guterding\corref{cor1}}
\ead{daniel.guterding@th-brandenburg.de}
\cortext[cor1]{Corresponding author.}

\address[inst1]{Technische Hochschule Brandenburg, Magdeburger Str.~50, 14770 Brandenburg an der Havel, Germany}
\address[inst2]{Department of Econometrics, Dokuz Eyl\"ul University, 35390 Izmir, T\"urkiye}

\begin{abstract}
The Nelson--Siegel--Svensson (NSS) interest rate curve model yields a separable nonlinear least-squares problem whose inner linear block is often ill-conditioned because the basis functions become nearly collinear. We analyze this instability via an exact orthogonal reparametrization of the design matrix. A thin QR decomposition produces orthogonal linear parameters for which, conditional on the nonlinear parameters, the Fisher information matrix is diagonal. We also derive a finite-horizon analytical orthogonalization: on $[0,T]$, the $4\times 4$ continuous Gram matrix has closed-form entries involving exponentials, logarithms, and the exponential integral $E_1$, yielding an explicit horizon-dependent orthogonal NSS basis. Together with Jacobian-rank and profile-likelihood arguments, this representation clarifies the degenerate manifold $\lambda_1=\lambda_2$, where the Svensson extension loses two degrees of freedom. Orthogonalization leaves the least-squares fit and uncertainty of the original linear parameters unchanged, but isolates the conditioning structure. When the decay parameters are estimated jointly, the full first-order covariance in orthogonal coordinates admits an explicit Schur-complement form. The approach also yields a scalar identifiability diagnostic through the QR element $R_{44}$ and separates model reduction from numerical instability. Synthetic experiments confirm that orthogonal parametrization eliminates correlations among the linear parameters and keeps their conditional uncertainty uniform. A daily U.S.\ Treasury study on a reduced fixed 9-tenor grid from 1981 to 2026 shows smoother orthogonal parameter series than classical NSS parameters while the moving QR basis remains nearly constant.
\end{abstract}


\end{frontmatter}

\section{Introduction}\label{sec:intro}

Estimating the term structure of interest rates is a classical inverse problem at the interface of numerical approximation, statistical calibration, and financial modeling. Two common families are flexible spline-based extraction methods, which enforce smoothness directly at the curve level~\citep{waggoner1997}, and low-dimensional parametric families such as the Nelson--Siegel (NS) and Nelson--Siegel--Svensson (NSS) models~\citep{nelson1987,svensson1994}. In central-bank and fixed-income applications, the parametric NS/NSS family remains particularly attractive because it provides compact, interpretable curve descriptions and can be recalibrated routinely on large historical data sets~\citep{bis2005,gurkaynak2007,diebold2013}.

From a computational point of view, however, the NSS model is more than a finance-specific parametrization. It is a small but delicate structured separable nonlinear least-squares problem: the model is nonlinear in the decay parameters~$\lambda$ and linear in the parameters~$\beta$, so the inner problem can be solved analytically while the outer problem remains non-convex. The associated design matrix can become nearly rank-deficient when the exponential scales $\lambda$ are close, so the linear parameters are formally separated but numerically unstable. The central contribution of this paper is to show that an exact orthogonal reparametrization of that inner linear block exposes the conditioning structure in a computationally useful way.

Existing calibration work reflects the same instability from several angles: fixed-shape or linearized variants reduce the problem to easier linear regressions~\citep{diebold2006,gauthier2012}; global and evolutionary search schemes try to mitigate false convergence in the nonlinear parameters~\citep{gimeno2009}; ridge or penalized formulations regularize unstable fits or steer the optimizer away from badly conditioned regions~\citep{annaert2013,banholzer2024}. We take a different but complementary approach. Rather than proposing another heuristic optimizer, we study what the standard QR/variable-projection geometry already reveals about conditioning, identifiability, regularization, and model reduction. This perspective is computational rather than macro-financial: the Treasury application later in the paper serves as a nontrivial calibration example, while the main contribution is the analysis of the underlying least-squares structure.

\subsection{Contribution}

In this paper, we taken an approach, which is complementary to the existing literature: rather than addressing the ill-conditioning only through penalization of the objective function (as in~\citep{banholzer2024}) or restriction of the parameter space (as in~\citep{depooter2007}), we analyze the problem through \emph{orthogonal reparametrization} of the basis functions.

Our contributions are:
\begin{enumerate}
\item \textbf{Orthogonal parametrization and stable inner solve.} We apply QR decomposition to the NSS design matrix $\Phi(\lambda)$, obtaining an orthonormal basis~$\Psi$ in which the Fisher information matrix for the linear parameters, conditional on~$\lambda$, is exactly diagonal. In this basis, all linear parameters are uncorrelated and have equal conditional uncertainty, regardless of the proximity of the decay parameters $\lambda_1$ and $\lambda_2$. Computationally, this isolates the ill-conditioning in the back-transformation $\beta = R^{-1}\gamma$ and turns the inner least-squares fit itself into the stable projection $\hat{\gamma} = \Psi^\top \hat{y}$.

\item \textbf{First-order estimator uncertainty with estimated decay parameters.} For the jointly estimated parameter vector $(\gamma,\lambda)$, we derive the local Gauss--Newton covariance and its explicit block form in orthogonal coordinates. The additional uncertainty beyond the conditional $\sigma^2 I_4$ term is governed by the projected nonlinear sensitivity $S = G^\top(I-\Psi\Psi^\top)G$, and the induced uncertainty of the classical parameters $\beta$ follows by a delta-method back-transformation through $R(\lambda)^{-1}$.

\item \textbf{Analytical continuous basis.} We derive a grid-independent orthogonalization on a finite maturity horizon $[0,T]$. All entries of the continuous $4\times 4$ Gram matrix admit closed forms in terms of exponentials, logarithms, and the exponential integral~$E_1$, so that the full analytical basis requires no auxiliary damping parameter. We also show that on $(0,\infty)$ no such fully parameter-free $4\times 4$ orthogonal basis exists in ordinary $L^2$, and that only the decaying $3\times 3$ subspace has a canonical limit.

\item \textbf{Structural identifiability analysis.} Using Taylor expansion, Jacobian rank conditions, and symmetry arguments, we characterize the identifiability of the NSS model. We show that the model is structurally identifiable when $\lambda_1 \neq \lambda_2$ and has a rank-2 deficiency at $\lambda_1 = \lambda_2$: the model reduces to four identifiable quantities ($\beta_1$, $\beta_2$, $\beta_3 + \beta_4$, $\lambda$).

\item \textbf{Profile likelihood analysis.} We compute profile likelihoods for both the standard and orthogonal parametrizations, demonstrating that, conditional on~$\lambda$, the orthogonal parameters have exactly quadratic (parabolic) profiles, consistent with well-identified parameters, while the standard parameters exhibit flat, correlated profiles in the near-degenerate regime. When profiling over all parameters including~$\lambda$, the profiles inherit nonlinearity from the $\lambda$-dependence of the QR decomposition.

\item \textbf{Identifiability diagnostic.} The diagonal element $|R_{44}|$ of the QR triangular factor serves as a scalar diagnostic for practical identifiability and conditioning: $|R_{44}| \to 0$ signals that the fourth basis function is nearly redundant, providing an automatic criterion for model reduction from Svensson to Nelson--Siegel.

\item \textbf{Regularization analysis.} We analyze the relationship between regularization in the standard and orthogonal bases, showing that penalizing $\|\gamma\|^2$ corresponds to uniform shrinkage across decorrelated components. While conceptually transparent, this uniform shrinkage is suboptimal in the degenerate regime, where standard ridge regression's differential shrinkage~\citep{hoerl1970} more effectively targets noise-amplifying directions. The orthogonal basis is most valuable as a diagnostic and model-selection tool (via $|R_{44}|$) rather than as a regularization framework.

\item \textbf{Analytical and nontrivial numerical examples.} We verify the conditioning and identifiability claims on synthetic yield curves covering both well-separated and nearly degenerate regimes. In addition, on daily US Treasury constant-maturity data fitted independently day by day, the orthogonal parameter series $\gamma_t$ are empirically smoother than the classical parameter series $\beta_t$. On a reduced 9-tenor grid chosen to avoid a historical 20Y tenor outage in the published data, the day-to-day basis rotation $\|\Psi(\lambda_{t-1})^\top \Psi(\lambda_t)-I\|_F$ remains extremely small, so the orthogonal parameter series are not only numerically well behaved within each day but also usable as a time series. As a supplementary robustness check, an exact dynamic-programming changepoint analysis of the monthly 4-dimensional histories finds essentially the same long-run regime map in $\beta$ and $\gamma$, but with materially lower within-regime dispersion for $\gamma$ at the same complexity.
\end{enumerate}

\subsection{Related work}

The related literature has several overlapping strands. A first strand concerns alternative representations of the yield curve itself. Spline and smoothing-spline approaches estimate discount, yield, or forward curves with roughness penalties and flexible knot structures~\citep{waggoner1997}. Central-bank technical studies also discuss practical production issues in yield-curve construction and the behavior of parametric specifications in institutional settings~\citep{bolder1999}. Our paper does not compete with those curve families; instead, it asks what can be learned by staying within the NSS family and changing coordinates inside the same model space.

A second strand concerns NSS-specific calibration strategies. The ill-conditioning of NSS fitting has been recognized since the original work of Nelson and Siegel~\citep{nelson1987}, who proposed a grid search over the nonlinear parameter to avoid non-convex optimization. De~Pooter~\citep{depooter2007} and Diebold and Li~\citep{diebold2006} fixed the decay parameter at predetermined values, sacrificing flexibility for stability. Gauthier and Simonato~\citep{gauthier2012} developed linearized NS/NSS variants that improve parameter stability and reduce computing effort, while Gimeno and Nave~\citep{gimeno2009} used a genetic algorithm to reduce the risk of false convergence in nonlinear term-structure estimation. Gilli et~al.~\citep{gilli2010} and Annaert et~al.~\citep{annaert2013} discussed multicollinearity more directly and proposed parameter-space restrictions or ridge regression~\citep{hoerl1970,belsley1980}.

The most relevant recent work is Banholzer et~al.~\citep{banholzer2024}, who provided the first rigorous mathematical analysis of the fitting problem. They formulated it as a separable nonlinear least-squares problem, proved smoothness of the reduced objective, characterized the rank-deficient regions, and proposed a penalty approach. Our work complements theirs by analyzing the same structured problem from the orthogonal-reparametrization perspective and adding identifiability, profile-likelihood, and model-reduction diagnostics. For direct yield fitting, Banholzer et~al.'s separation between the nonlinear decay parameters~$\lambda$ and the inner linear parameters~$\beta$ is the appropriate mathematical formulation. In practical calibration stacks, however, NSS is often embedded in broader nonlinear pricing workflows rather than treated as an explicit yield-regression inner solve. In that setting, the restartability and warm-start behavior of the orthogonal $\gamma$-based parametrization become operationally relevant, a point illustrated by the long-history calibration study in Section~\ref{sec:timeseries}.

The separable structure of the NSS fitting problem also connects to the broader variable-projection literature. Beyond the foundational papers of Golub and Pereyra~\citep{golub1973,golub2003}, classical analyses by Ruhe and Wedin~\citep{ruhe1980} and implementation work by Kaufman~\citep{kaufman1975} and O'Leary and Rust~\citep{oleary2013} show that QR-based elimination of the linear parameters is a standard computational device for separable nonlinear least-squares problems. Our contribution is not the orthogonalization itself---which is implicit in any QR-based VARPRO implementation---but rather the systematic analysis of what the orthogonal parametrization reveals about identifiability, the $R_{44}$ diagnostic for model selection, the finite-horizon analytical basis, and the profile likelihood comparison specific to the NSS model. The concept of parameter orthogonality also has a long history in statistics, notably the work of Cox and Reid~\citep{cox1987} on adjusted profile likelihoods with orthogonal parameters. Our orthogonalization achieves conditional parameter independence within the linear-parameter subblock in the sense of Cox and Reid~\citep{cox1987}: the $4 \times 4$ Fisher information for $\gamma$ given $\lambda$ is diagonal. However, the full $\gamma$--$\lambda$ orthogonality required for Cox--Reid adjusted profile likelihoods does not hold, as discussed in Remark~\ref{rem:fullFisher}.

Related inverse problems in quantitative finance display a similar pattern of ill-conditioned linear systems but lead to different stabilization strategies. In the context of arbitrage-free interpolation of option prices and implied volatilities, \citet{guterding2023} derives a discretized pricing kernel whose matrix is generically ill-conditioned and stabilizes the problem by an SVD-based sparse representation with $L^1$ regularization. That work addresses a different task---stable recovery in an inverse problem with no fixed parametric factor model---but it reinforces the broader point that ill-conditioning is a structural issue in financial curve reconstruction problems, not an implementation detail. Our QR-based orthogonalization is complementary: it preserves the NSS model family exactly, yields interpretable orthogonal parameters, and produces the $R_{44}$ identifiability diagnostic.

Several open-source implementations of NSS fitting are available, including the Python package \path{nelson_siegel_svensson}, dynamic Nelson--Siegel--Svensson Kalman-filter implementations, the R package \texttt{YieldCurve}, and the general-purpose VARPRO implementations \texttt{varpro2}~\citep{oleary2013} (MATLAB) and \texttt{varpro} (Rust). Our contribution is not a competing implementation but rather a reparametrization and diagnostic framework. The proposed QR factorization is part of the inner linear solve of a VARPRO-style yield-regression implementation. When decay-parameter estimates are supplied by an existing pipeline, those quantities can also be used as a post-fit diagnostic.

\section{The Nelson--Siegel--Svensson Model}\label{sec:model}

\subsection{Model specification}

Given maturities $0 < \tau_1 < \cdots < \tau_m \leq T$ and observed yields $\hat{y} = (\hat{y}_1, \ldots, \hat{y}_m)^\top$, the Svensson model curve is
\begin{equation}\label{eq:nss}
y_{\lambda,\beta}(\tau) = \beta_1 + \beta_2 \frac{1 - e^{-\lambda_1\tau}}{\lambda_1\tau} + \beta_3\left(\frac{1 - e^{-\lambda_1\tau}}{\lambda_1\tau} - e^{-\lambda_1\tau}\right) + \beta_4\left(\frac{1 - e^{-\lambda_2\tau}}{\lambda_2\tau} - e^{-\lambda_2\tau}\right),
\end{equation}
with linear parameters $\beta = (\beta_1, \beta_2, \beta_3, \beta_4)^\top \in \mathbb{R}^4$ and nonlinear decay parameters $\lambda = (\lambda_1, \lambda_2)^\top \in \mathbb{R}^2_+$. The Nelson--Siegel model is obtained by setting $\beta_4 = 0$ (equivalently, dropping the fourth basis function).

In matrix form, $y(\tau) = \Phi(\lambda)\beta$ where the design matrix $\Phi(\lambda) \in \mathbb{R}^{m \times 4}$ has entries $\Phi_{ij} = \phi_j(\lambda; \tau_i)$ for the four basis functions.

\subsection{Separable structure and the reduced problem}

Following~\citep{banholzer2024}, the fitting problem
\begin{equation}
\min_{\lambda \in \Lambda,\, \beta \in \mathbb{R}^4} F(\lambda, \beta) := \|\Phi(\lambda)\beta - \hat{y}\|_2^2
\end{equation}
is a separable nonlinear least-squares problem~\citep{golub1973,golub2003}. For fixed $\lambda$, the optimal $\beta^*(\lambda) = \Phi(\lambda)^\dagger \hat{y}$, and the reduced objective is
\begin{equation}\label{eq:reduced}
H(\lambda) = \|\Phi(\lambda)\Phi(\lambda)^\dagger\hat{y} - \hat{y}\|_2^2 = \hat{y}^\top(I - P_\lambda)\hat{y},
\end{equation}
where $P_\lambda = \Phi(\lambda)\Phi(\lambda)^\dagger$ is the projection onto the column space of $\Phi(\lambda)$.

\section{Orthogonal Reparametrization}\label{sec:ortho}

\subsection{QR decomposition of the design matrix}

For fixed $\lambda$, the design matrix $\Phi(\lambda) \in \mathbb{R}^{m \times 4}$ admits a thin QR decomposition:
\begin{equation}\label{eq:qr}
\Phi(\lambda) = \Psi(\lambda) R(\lambda),
\end{equation}
where $\Psi(\lambda) \in \mathbb{R}^{m \times 4}$ has orthonormal columns ($\Psi^\top\Psi = I_4$) and $R(\lambda) \in \mathbb{R}^{4 \times 4}$ is upper triangular. We enforce the convention of positive diagonal entries by flipping signs of columns of~$\Psi$ and corresponding rows of~$R$ when $R_{jj} < 0$; see the accompanying code for details. We use unpivoted QR throughout. For the standard NSS ordering in Eq.~\eqref{eq:nss}, this places the relevant near-redundant direction last: columns 1--3 span the Nelson--Siegel subspace, and column~4 is the Svensson extension that collapses onto $\mathrm{span}\{\phi_1, \phi_2, \phi_3\}$ as $\lambda_2 \to \lambda_1$. Consequently, unpivoted Householder QR reveals the degeneracy of the Svensson extension in $R_{44}$ without permutation. This statement depends on the NSS column ordering; for reordered bases or generalizations with more basis functions, column-pivoted QR~\citep{stewart1998,duersch2025} would automatically identify the near-redundant directions regardless of column ordering, and is available at nearly the cost of unpivoted QR~\citep{duersch2025}.

\begin{remark}
The columns of $\Psi$ form an orthonormal basis for the same column space as $\Phi$. The yield curve model becomes $y(\tau) = \Psi(\lambda)\gamma$ where $\gamma = R(\lambda)\beta$ are the \emph{orthogonal parameters}.
\end{remark}

\begin{remark}\label{rem:notation}
Unless explicitly decorated by a subscript~$T$, the symbols $\Psi$ and $R$ refer to the discrete QR factors of the design matrix on the observed maturity grid. The continuous finite-horizon basis is denoted by $\Psi_T$ and its triangular factor by~$R_T$.
\end{remark}

\subsection{Continuous analytical orthogonalization on a finite horizon}

If one seeks a grid-independent reference basis, the corresponding continuous construction is the $L^2(0,T)$ inner product
\begin{equation}
\langle f, g \rangle_T = \int_0^T f(\tau)g(\tau)\,d\tau,
\end{equation}
where $T > 0$ denotes a fixed maturity horizon, for example the economically relevant maximum maturity used in a given application. Let $\varphi(\tau) = (\phi_1(\tau), \phi_2(\tau), \phi_3(\tau), \phi_4(\tau))^\top$ and define the continuous Gram matrix
\begin{equation}
G_T(\lambda) = \bigl(\langle \phi_i, \phi_j\rangle_T\bigr)_{i,j=1}^4.
\end{equation}
Since all four NSS basis functions lie in $L^2(0,T)$, the matrix $G_T(\lambda)$ is symmetric positive definite for $\lambda_1 \neq \lambda_2$, and its Cholesky factorization
\begin{equation}
G_T(\lambda) = L_T(\lambda)L_T(\lambda)^\top
\end{equation}
defines a continuous orthonormal basis
\begin{equation}
\psi_T(\tau) = L_T(\lambda)^{-1}\varphi(\tau).
\end{equation}
Equivalently, when these functions are evaluated on any discrete grid, one has $\Phi = \Psi_T R_T$ with $R_T = L_T^\top$, directly paralleling the discrete QR factorization.

\begin{proposition}[Finite-horizon analytical basis]\label{prop:finiteT}
For any $T>0$ and $\lambda_1 \neq \lambda_2$, the continuous Gram matrix $G_T(\lambda)$ is symmetric positive definite, and all of its entries admit closed forms in terms of exponentials, logarithms, and the exponential integral $E_1$. Consequently, the Cholesky factorization $G_T = L_TL_T^\top$ yields a full $4\times4$ analytical orthogonal NSS basis on $[0,T]$ via $\psi_T(\tau)=L_T^{-1}\varphi(\tau)$, without any auxiliary damping parameter (as it would be necessary for $T \to \infty$). The explicit formulas for the Gram-matrix entries are collected in \ref{app:finiteT-derivation}.
\end{proposition}

\begin{remark}[Interpretation of the finite-horizon basis]\label{rem:finitebasis}
The basis $\Psi_T$ is horizon-dependent: changing $T$ changes the orthogonal directions, the corresponding triangular factor $R_T$, and the continuous analogue of the $R_{44}$ diagnostic. We therefore interpret $\Psi_T$ as a grid-independent analytical reference basis on a chosen economic horizon, not as a replacement for the discrete QR basis that exactly diagonalizes the Fisher information of a given OLS or WLS problem on an observed maturity grid.
\end{remark}

\begin{remark}[Why there is no full parameter-free basis on $(0,\infty)$]\label{rem:infinitebasis}
On $(0,\infty)$ in the ordinary $L^2$ inner product, the level mode $\phi_1(\tau)=1$ is not square-integrable, whereas $\phi_2,\phi_3,\phi_4 \in L^2(0,\infty)$ because they decay as $O(\tau^{-1})$. Consequently, no full parameter-free $4\times4$ orthonormal NSS basis exists on $(0,\infty)$. What does exist is the canonical $3\times3$ Gram matrix of the decaying subspace $\mathrm{span}\{\phi_2,\phi_3,\phi_4\}$, which yields an infinite-horizon reference value for the lower-right block and, in particular, for $R_{44}$.
\end{remark}

\subsection{Properties of the orthogonal parametrization}

Since $\Psi^\top\Psi = I$, all singular values of $\Psi$ equal~1, and the condition number $\kappa(\Psi) = 1$ for any $\lambda \in \mathbb{R}^2_+$. In finite-precision arithmetic, the computed $\hat{\Psi}$ satisfies $\|\hat{\Psi}^\top\hat{\Psi} - I\|_F = O(\kappa(\Phi)\,\varepsilon_{\mathrm{mach}})$, where $\varepsilon_{\mathrm{mach}}$ is the unit round-off (for double precision, $\varepsilon_{\mathrm{mach}}\approx 2.22 \times 10^{-16}$)~\citep{higham2002}. For the condition numbers encountered in realistic examples ($\kappa(\Phi) \leq 10^4$), the orthogonality loss is $O(10^{-12})$, negligible in double precision. The broader condition-number map in Fig.~\ref{fig:cond} includes points much closer to singular sets, where $\kappa(\Phi)$ is larger. In orthogonal parametrization, the inner linear subproblem is perfectly conditioned by construction.

\begin{proposition}[Conditional diagonal Fisher information]\label{prop:fisher}
Under the model $\hat{y} = \Psi(\lambda)\gamma + \varepsilon$ with $\varepsilon \sim \mathcal{N}(0, \sigma^2 I)$ and $\lambda$ treated as known, the Fisher information matrix for $\gamma$ is
\begin{equation}
\mathcal{I}_{\gamma|\lambda} = \frac{1}{\sigma^2}\Psi^\top\Psi = \frac{1}{\sigma^2}I_4.
\end{equation}
Consequently, $\mathrm{Cov}(\hat{\gamma}\mid\lambda) = \sigma^2 I_4$ and all linear parameters have equal conditional variance $\sigma^2$ and zero correlation.
\end{proposition}

\begin{remark}\label{rem:fullFisher}
The full $6 \times 6$ Fisher information matrix for $(\gamma, \lambda)$ is generally \emph{not} diagonal. The off-diagonal blocks coupling $\gamma$ and $\lambda$ persist after orthogonalization because $\gamma = R(\lambda)\beta$ depends nonlinearly on~$\lambda$. The diagonal structure applies only to the $4 \times 4$ linear-parameter subblock, conditional on~$\lambda$.
\end{remark}

\begin{corollary}[Conditional invariance to $\lambda$ proximity]
For any $\lambda \in \mathbb{R}^2_+$, the conditional uncertainty of $\hat{\gamma}$ is independent of $\lambda$. In particular, $\mathrm{Cov}(\hat{\gamma}\mid\lambda) = \sigma^2 I_4$ does not degrade as $\lambda_1 \to \lambda_2$, in sharp contrast to $\mathrm{Cov}(\hat{\beta}\mid\lambda) = \sigma^2(\Phi^\top\Phi)^{-1}$ which diverges. The marginal uncertainty of $\hat{\gamma}$, accounting for estimation uncertainty in $\lambda$, will generally differ from $\sigma^2 I_4$.
\end{corollary}

\subsection{First-order full covariance with estimated decay parameters}\label{sec:fullcov}

The conditional result above is exact for fixed~$\lambda$. For joint estimation of $(\gamma,\lambda)$, the relevant local object is the Gauss--Newton / observed-Fisher approximation to the full covariance.

\begin{proposition}[First-order covariance in orthogonal coordinates]\label{prop:fullcov}
Let $\eta = (\gamma,\lambda) \in \mathbb{R}^4 \times \mathbb{R}^2_+$ and define the model map
\[
f(\eta) = \Psi(\lambda)\gamma \in \mathbb{R}^m.
\]
At a locally identifiable point $\eta_0 = (\gamma_0,\lambda_0)$ where the map $\lambda \mapsto \Psi(\lambda)$ is differentiable, let
\[
K = \frac{\partial f}{\partial \eta}(\eta_0) = [\Psi,\ G], \qquad
G = [g_1, g_2], \qquad
g_j = \partial_{\lambda_j}(\Psi(\lambda)\gamma)\big|_{\eta_0}.
\]
Assume that $K$ has full column rank. Under the local linearization
\[
f(\eta_0+\delta) \approx f(\eta_0) + K\delta
\]
with homoscedastic Gaussian errors of variance $\sigma^2$, the joint least-squares estimator $\hat{\eta} = (\hat{\gamma},\hat{\lambda})$ has the observed-Fisher covariance approximation
\[
\mathrm{Cov}(\hat{\eta}) \approx \sigma^2 (K^\top K)^{-1}.
\]
Writing
\[
C = \Psi^\top G \in \mathbb{R}^{4\times 2}, \qquad
M = G^\top G \in \mathbb{R}^{2\times 2}, \qquad
S = M - C^\top C = G^\top (I - \Psi\Psi^\top) G,
\]
one has
\[
K^\top K =
\begin{pmatrix}
I_4 & C \\
C^\top & M
\end{pmatrix},
\]
and, because $\Psi^\top\Psi = I_4$, the matrix $K^\top K$ is nonsingular if and only if $S$ is nonsingular. In that case,
\[
(K^\top K)^{-1}
=
\begin{pmatrix}
I_4 + C S^{-1} C^\top & -C S^{-1} \\
-S^{-1} C^\top & S^{-1}
\end{pmatrix}.
\]
Consequently,
\[
\mathrm{Cov}(\hat{\lambda}) \approx \sigma^2 S^{-1}, \qquad
\mathrm{Cov}(\hat{\gamma}) \approx \sigma^2\bigl(I_4 + C S^{-1} C^\top\bigr),
\]
\[
\mathrm{Cov}(\hat{\gamma},\hat{\lambda}) \approx -\sigma^2 C S^{-1}.
\]
\end{proposition}

\begin{proof}
In the linearized model, the parameter increment $\delta$ is estimated by ordinary least squares:
\[
\hat{\delta} = \arg\min_\delta \|\varepsilon - K\delta\|_2^2.
\]
Since $K$ has full column rank, the normal equations give
\[
\hat{\delta} = (K^\top K)^{-1}K^\top\varepsilon,
\]
hence $\mathrm{Cov}(\hat{\eta}) \approx \mathrm{Cov}(\hat{\delta}) = \sigma^2(K^\top K)^{-1}$. Since $\Psi^\top\Psi = I_4$, the stated block form of $K^\top K$ follows immediately. The inverse is the standard block-inverse formula with upper-left block $I_4$ and Schur complement $S = M - C^\top C$, and $K^\top K$ is nonsingular if and only if $S$ is.
\end{proof}

\begin{remark}[Interpretation of the Schur complement]\label{rem:schur}
The matrix $S = G^\top(I-\Psi\Psi^\top)G$ measures the part of the $\lambda$-sensitivity that cannot be absorbed by retuning $\gamma$ inside the current four-dimensional model span. Thus $\lambda$ is informed only by changes in the fitted curve that leave that span. If $S$ becomes small, the decay parameters are weakly identified, and the additional uncertainty from estimating $\lambda$ propagates back into $\gamma$ through the coupling matrix~$C$.
\end{remark}

\begin{corollary}[Delta-method covariance for recovered classical parameters]\label{cor:beta-cov}
Let $\dot{R}_j = \partial_{\lambda_j}R(\lambda)$ at the same point $\eta_0$. Since $\beta = R(\lambda)^{-1}\gamma$, its differential is
\[
d\beta
=
R^{-1}d\gamma
- R^{-1}\dot{R}_1\beta\, d\lambda_1
- R^{-1}\dot{R}_2\beta\, d\lambda_2.
\]
Hence the Jacobian of the back-transformation is
\[
D_\eta \beta
=
\bigl[
R^{-1},\,
-R^{-1}\dot{R}_1\beta,\,
-R^{-1}\dot{R}_2\beta
\bigr],
\]
and the recovered classical parameters satisfy the first-order covariance formula
\[
\mathrm{Cov}(\hat{\beta})
\approx
D_\eta\beta\,
\mathrm{Cov}(\hat{\eta})\,
(D_\eta\beta)^\top.
\]
If $\lambda$ is treated as known, this reduces to the familiar conditional covariance
\[
\mathrm{Cov}(\hat{\beta}\mid\lambda)
=
\sigma^2 R^{-1}R^{-\top}
=
\sigma^2(\Phi^\top\Phi)^{-1}.
\]
\end{corollary}

\begin{proof}
Differentiate $\beta = R^{-1}\gamma$ and use $d(R^{-1}) = -R^{-1}(dR)R^{-1}$. The covariance statement then follows from the delta method applied to Proposition~\ref{prop:fullcov}.
\end{proof}

\subsection{Parameter recovery and interpretation}

The original parameters are recovered via $\beta = R^{-1}\gamma$. The economic interpretation of the orthogonal parameters $\gamma$ is as projections of the yield curve onto orthogonal ``modes'': $\gamma_j = \psi_j^\top \hat{y}$, representing the component of the observed curve in the $j$-th orthogonal direction.

\subsection{Estimation in \texorpdfstring{$\gamma$}{gamma} coordinates}\label{sec:fit-gamma}

Although the least-squares solution is the same in both parametrizations---$\hat{\beta} = \Phi^\dagger\hat{y}$ and $\hat{\beta} = R^{-1}(\Psi^\top\hat{y})$ give the \emph{same} $\hat{\beta}$---the computational path differs. Estimation in the standard basis requires solving the normal equations $(\Phi^\top\Phi)\hat{\beta} = \Phi^\top\hat{y}$ or the triangular system $R\hat{\beta} = \Psi^\top\hat{y}$; both are ill-conditioned when $\kappa(\Phi)$ is large. In the orthogonal basis, the inner fit reduces to the stable projection
\begin{equation}\label{eq:gamma-fit}
\hat{\gamma}_j = \psi_j^\top \hat{y}, \quad j = 1,\ldots,4,
\end{equation}
with $\kappa(\Psi) = 1$ by construction. The ill-conditioning is encapsulated in a single step: the recovery $\hat{\beta} = R^{-1}\hat{\gamma}$, which involves inverting the triangular factor~$R$ with small diagonal entry~$R_{44}$. This step is unnecessary when inference or prediction is carried out directly in the $\gamma$-parametrization.

Consequently, estimation and prediction are most naturally carried out in the orthogonal coordinates. One computes $\hat{\gamma} = \Psi^\top\hat{y}$, which is numerically stable because $\kappa(\Psi)=1$, and forms the fitted curve as $\hat{y}(\tau) = \Psi(\lambda)\hat{\gamma}$, identical to $\Phi(\lambda)\hat{\beta}$ without requiring explicit recovery of~$\hat{\beta}$. In this representation, model selection between Nelson--Siegel and Nelson--Siegel--Svensson amounts to excluding or including~$\gamma_4$; because $\Psi$ is orthonormal, the estimates $\hat{\gamma}_1$, $\hat{\gamma}_2$, and $\hat{\gamma}_3$ are unchanged by that choice, unlike their counterparts in the $\beta$-basis. When the classical economic interpretation is needed, the original parameters may be recovered from $\hat{\beta} = R^{-1}\hat{\gamma}$, but this back-transformation is reliable only when $|R_{44}|$ is sufficiently large.

Conditional on $\lambda$, the profile likelihoods of the $\gamma_j$ are exactly quadratic (Section~\ref{sec:profile}). The diagonal entry $R_{44}$ becomes a scalar identifiability diagnostic (Section~\ref{sec:diagnostic}), model selection reduces to adding or dropping~$\gamma_4$ without refitting the remaining orthogonal parameters, and the same basis provides the natural setting for the regularization analysis of Section~\ref{sec:regularization}.

\begin{algorithm}[tb]                                                                                                                                          
  \caption{Inner step: orthogonal NSS solve with $R_{44}$ diagnostic for given $\lambda$}
  \label{alg:ortho}                                        
  \begin{algorithmic}[1]                      
  \Require Maturities $\tau$, observed yields $\hat{y}$, current iterate $\lambda = (\lambda_1, \lambda_2)$ of the outer VARPRO loop, threshold $\sigma/\delta$
  \State Compute design matrix $\Phi(\lambda) \in \mathbb{R}^{m \times 4}$ \Comment{Eq.~\eqref{eq:nss}}                                                          
  \State Thin QR: $\Phi = \Psi R$ with $R_{jj} > 0$ \Comment{Eq.~\eqref{eq:qr}}
  \State Orthogonal fit: $\hat{\gamma} = \Psi^\top \hat{y}$ \Comment{Always stable, $\kappa = 1$}
  \State Set $p \leftarrow 3$ if $|R_{44}| < \sigma/\delta$, else $p \leftarrow 4$ \Comment{Identifiability diagnostic}                                          
  \State Fitted curve: $f = \Psi_{:,1:p}\,\hat{\gamma}_{1:p}$ \Comment{Nelson--Siegel if $p=3$, Svensson if $p=4$}                                               
  \If{economic parameters $\beta$ needed}                                                
    \If{$p = 4$}
      \State $\hat{\beta} = R^{-1}\hat{\gamma}$ \Comment{Full recovery}
    \Else                                                      
      \State $\hat{\beta}_{1:3} = R_3^{-1}\hat{\gamma}_{1:3}$, \; $\hat{\beta}_4 = 0$ \Comment{$R_3$: leading $3\times 3$ subblock of $R$}
    \EndIf
  \EndIf
  \State \Return $\hat{\gamma}_{1:p}$, $|R_{44}|$, $\kappa(\Phi)$, (optionally $\hat{\beta}$)      
  \end{algorithmic}                                               
  \end{algorithm}

\section{Local Structural Identifiability}\label{sec:structural}

\subsection{Jacobian rank condition}

The model is locally structurally identifiable at a parameter point $\theta = (\beta, \lambda)$ if the sensitivity matrix
\begin{equation}
J(\theta) = \begin{pmatrix} \frac{\partial y(\tau_1)}{\partial\theta} \\ \vdots \\ \frac{\partial y(\tau_m)}{\partial\theta} \end{pmatrix} \in \mathbb{R}^{m \times 6}
\end{equation}
has full column rank~6~\citep{rothenberg1971}. We use this standard local Jacobian-rank criterion while noting the broader distinction between local and global identifiability and the related notion of parameter redundancy in nonlinear parametrizations~\citep{ljung1994,catchpole1997}.

\begin{proposition}[Structural identifiability]\label{prop:ident}
Let $\tau_1, \ldots, \tau_m$ with $m \geq 6$ be maturities such that no subset of five maturities is a zero set of any nonzero linear combination of the NSS sensitivity functions. Then:
\begin{enumerate}
\item[(a)] If $\lambda_1 \neq \lambda_2$ (both $> 0$) and $\beta$ is generic in the sense that $\beta_3\beta_4 \neq 0$, then $\mathrm{rank}(J) = 6$ and the model is locally structurally identifiable.
\item[(b)] If $\lambda_1 = \lambda_2 = \lambda$, then $\phi_3(\lambda;\tau) \equiv \phi_4(\lambda;\tau)$ for all $\tau > 0$, and the six sensitivity functions $\partial y/\partial\theta_i$ all lie in the four-dimensional function space $\mathrm{span}\{1,\, \phi_2(\lambda;\tau),\, e^{-\lambda\tau},\, \tau e^{-\lambda\tau}\}$. Hence $\mathrm{rank}(J) \leq 4$. For generic $\beta$ (with $\beta_4 \neq 0$), the rank is exactly~4, with a two-dimensional null space. The first null direction is the trade-off $\beta_3 \leftrightarrow \beta_4$ at constant $\beta_3 + \beta_4$; the second couples an infinitesimal split $\lambda_1 - \lambda_2$ to a compensating redistribution of $\beta_3$ and $\beta_4$. As $\lambda_1 \rightarrow \lambda_2$ the model becomes parameter-redundant of deficiency 2. Therefore, the degenerate model has exactly four identifiable quantities: $\beta_1$, $\beta_2$, $\beta_3 + \beta_4$, and $\lambda$.
\end{enumerate}
\end{proposition}

\begin{proof}[Proof of part (b)]
First, $\phi_3(\lambda;\tau) = \phi_4(\lambda;\tau)$ when $\lambda_1 = \lambda_2 = \lambda$, collapsing columns~3 and~4 of~$\Phi$. Second, the identity $\partial\phi_2/\partial\lambda = \lambda^{-1}(e^{-\lambda\tau} - \phi_2)$ shows that the $\lambda$-derivative of~$\phi_2$ lies in $\mathrm{span}\{\phi_2, e^{-\lambda\tau}\}$, so the $\lambda$-sensitivities add only $\tau e^{-\lambda\tau}$ as a genuinely new function. The column space of~$J$ is therefore at most four-dimensional.
\end{proof}

\begin{proof}[Proof of part (a)]
Set $g_i(\tau) = \tau \, \partial y(\tau)/\partial\theta_i$. Since $\tau > 0$ on $(0,\infty)$, the six sensitivity functions are linearly independent if and only if the six functions $g_i$ are linearly independent. A direct calculation gives
\begin{align*}
g_{\beta_1}(\tau) &= \tau, \\
g_{\beta_2}(\tau) &= \lambda_1^{-1}(1 - e^{-\lambda_1 \tau}), \\
g_{\beta_3}(\tau) &= \lambda_1^{-1}(1 - e^{-\lambda_1 \tau}) - \tau e^{-\lambda_1 \tau}, \\
g_{\beta_4}(\tau) &= \lambda_2^{-1}(1 - e^{-\lambda_2 \tau}) - \tau e^{-\lambda_2 \tau}, \\
g_{\lambda_1}(\tau) &= -\frac{\beta_2 + \beta_3}{\lambda_1^2}
+ \frac{\beta_2 + \beta_3}{\lambda_1^2} e^{-\lambda_1 \tau}
+ \frac{\beta_2 + \beta_3}{\lambda_1} \tau e^{-\lambda_1 \tau}
+ \beta_3 \tau^2 e^{-\lambda_1 \tau}, \\
g_{\lambda_2}(\tau) &= -\frac{\beta_4}{\lambda_2^2}
+ \frac{\beta_4}{\lambda_2^2} e^{-\lambda_2 \tau}
+ \frac{\beta_4}{\lambda_2} \tau e^{-\lambda_2 \tau}
+ \beta_4 \tau^2 e^{-\lambda_2 \tau}.
\end{align*}
Hence each $g_i$ lies in the span of
\[
\mathcal{B} = \{1, \tau, e^{-\lambda_1 \tau}, \tau e^{-\lambda_1 \tau}, \tau^2 e^{-\lambda_1 \tau}, e^{-\lambda_2 \tau}, \tau e^{-\lambda_2 \tau}, \tau^2 e^{-\lambda_2 \tau}\}.
\]
For $\lambda_1 \neq \lambda_2$, the eight functions in $\mathcal{B}$ are linearly independent: they form the standard solution basis of the constant-coefficient ODE, where $D = d/d\tau$,
\[
D^2 (D + \lambda_1)^3 (D + \lambda_2)^3 f(\tau) = 0.
\]
Now suppose $\sum_{i=1}^6 c_i g_i(\tau) \equiv 0$. Comparing coefficients in the basis $\mathcal{B}$ yields
\[
c_{\lambda_1} \beta_3 = 0,
\qquad
c_{\lambda_2} \beta_4 = 0
\]
from the $\tau^2 e^{-\lambda_1 \tau}$ and $\tau^2 e^{-\lambda_2 \tau}$ terms. Under the generic condition $\beta_3 \beta_4 \neq 0$, this forces $c_{\lambda_1} = c_{\lambda_2} = 0$. The coefficients of $\tau e^{-\lambda_2 \tau}$ and $\tau e^{-\lambda_1 \tau}$ then give $c_{\beta_4} = c_{\beta_3} = 0$, the coefficient of $e^{-\lambda_1 \tau}$ gives $c_{\beta_2} = 0$, and the coefficient of $\tau$ gives $c_{\beta_1} = 0$. Therefore $c_i = 0$ for all $i$, so the six sensitivities are linearly independent on $(0,\infty)$. Evaluating them at non-degenerate maturities preserves rank, hence $\mathrm{rank}(J) = 6$.
\end{proof}

\subsection{Taylor expansion analysis}

Expanding $y(\tau)$ in powers of $\tau$ around $\tau = 0$:
\begin{equation}
y(\tau) = (\beta_1 + \beta_2) + \frac{1}{2}\bigl[\lambda_1(\beta_3 - \beta_2) + \lambda_2\beta_4\bigr]\tau + O(\tau^2).
\end{equation}
The short rate $y(0) = \beta_1 + \beta_2$ and the long rate $\lim_{\tau\to\infty} y(\tau) = \beta_1$ are always identifiable. The slope at $\tau = 0$ identifies the aggregate quantity $\lambda_1(\beta_3 - \beta_2) + \lambda_2\beta_4$. When $\lambda_1 = \lambda_2 = \lambda$, this becomes $\lambda(\beta_3 - \beta_2 + \beta_4)$, and the individual contributions of $\beta_3$ and $\beta_4$ are no longer separable, consistent with the rank-2 deficiency established in Proposition~\ref{prop:ident}.

\section{Profile Likelihood Analysis}\label{sec:profile}                                           
Under Gaussian errors, the log-likelihood is $\ell(\theta) = -\frac{m}{2}\log(2\pi\sigma^2) - \frac{1}{2\sigma^2}\|\hat{y} - f(\theta)\|^2$, where $f(\theta) = \Phi(\lambda)\beta$ is the model-predicted yield curve. The \emph{profile negative log-likelihood} for parameter $\theta_j$, shifted so that its minimum at the MLE is zero, is obtained by minimizing $-\ell$ over all other parameters~\citep{raue2009}:
\begin{equation}                                                                                   \label{eq:NLL}                                                                                     
\Delta\mathrm{NLL}(\theta_j)\;=\; \min_{\theta_{-j}}\bigl[-\ell(\theta)\bigr] \;-\; \bigl[-\ell(\hat\theta)\bigr]\;=\; \frac{1}{2\sigma^2}\Bigl[\min_{\theta_{-j}}\|\hat{y}-f(\theta)\|^2 \;-\; \|\hat{y}-f(\hat\theta)\|^2\Bigr].                                              
\end{equation}
Up to a factor of two, this is the Wilks profile deviance $\chi^2_{\mathrm{prof}}(\theta_j) = -2\log\!\bigl[L_{\mathrm{prof}}(\theta_j)/L_{\max}\bigr] = 2\,\Delta\mathrm{NLL}(\theta_j)$, so a $95\%$ confidence interval for $\theta_j$ is the set $\{\theta_j : \Delta\mathrm{NLL}(\theta_j) \le \tfrac{1}{2}\chi^2_{1,0.95} \approx 1.92\}$.   
\subsection{Profiles in the standard parametrization}
For the linear parameters $\beta_j$ (with $\lambda$ fixed), profiling out the remaining coefficients $\beta_{-j}$ yields the exact quadratic                   
\begin{equation}
\Delta\mathrm{NLL}(\beta_j \mid \lambda)\;=\; \frac{(\beta_j - \hat\beta_j)^2}{2\,\sigma_{\mathrm{prof},j}^2},\qquad \sigma_{\mathrm{prof},j}^2 \;=\; \frac{\sigma^2}{\|\phi_j^\perp\|^2},
\end{equation}                                                                                     where $\phi_j^\perp = Q_{-j}\phi_j$ is the component of $\phi_j$ orthogonal to the remaining columns of~$\Phi$ and $Q_{-j}$ projects onto $\mathrm{span}(\Phi_{-j})^\perp$. This profile variance equals the Fisher-covariance diagonal exactly, $\sigma^2_{\mathrm{prof},j} = [\Sigma_\beta]_{jj}$, which admits the standard variance-inflation decomposition
\begin{equation}                                                                                   \sigma^2_{\mathrm{prof},j}\;=\; [\Sigma_\beta]_{jj}\;=\; \frac{\sigma^2}{\|\phi_j\|^2}\cdot\frac{1}{1-\rho_j^2},
\end{equation}                                                            
where $\rho_j$ is the multiple correlation between column $\phi_j$ and the remaining columns of $\Phi$ (i.e.\ $\rho_j^2 = R^2$ of regressing $\phi_j$ on $\Phi_{-j}$), and $1/(1-\rho_j^2)$ is the variance inflation factor for column~$j$. As columns become collinear ($\rho_j \to 1$), $\|\phi_j^\perp\| \to 0$ and $\sigma^2_{\mathrm{prof},j} \to \infty$, so the profile flattens---indicating practical non-identifiability~\citep{raue2009}.
\subsection{Profiles in the orthogonal parametrization}
For $\gamma_j$, conditional on fixed $\lambda$, since $\mathrm{Cov}(\hat{\gamma}\mid\lambda) = \sigma^2 I$, the profile is exactly:
\begin{equation}                                                                                   \Delta\mathrm{NLL}(\gamma_j \mid \lambda)\;=\; \frac{(\gamma_j - \hat\gamma_j)^2}{2\,\sigma^2}.
\end{equation}
a perfect parabola in the log-likelihood with curvature $1/\sigma^2$ independent of $j$ and $\lambda$. This is the characteristic profile of a well-identified parameter in the conditional (inner) problem.
\begin{remark}[Conditional vs.\ full profiles]
The exact quadratic $\Delta\mathrm{NLL}$ result holds strictly for $\gamma$ given fixed $\lambda$. When profiling $\gamma_j$ over all parameters \emph{including} $\lambda_1$ and $\lambda_2$, the profile inherits nonlinearity because $\gamma = R(\lambda)\beta$ couples the linear and nonlinear parameters through the $\lambda$-dependent QR decomposition. The full profile over all six parameters must be interpreted separately from the conditional profiles presented here.
\end{remark}

To assess the practical importance of the conditional vs.\ full distinction, we computed full profiles (optimizing over all parameters including $\lambda$) for $\beta_4$ and $\gamma_4$ in the near-degenerate regime. The full profiles are wider and non-quadratic, reflecting the additional uncertainty from $\lambda$ estimation~\citep{simpson2023}. Proposition~\ref{prop:fullcov} gives the corresponding covariance statement: once $\lambda$ is estimated, the orthogonal coordinates acquire the positive-semidefinite inflation term $\sigma^2 C S^{-1}C^\top$, and the recovered classical parameters receive an additional delta-method contribution through the $\lambda$-dependence of~$R$. The orthogonal parametrization therefore provides no generic advantage for these full profiles: the practical identifiability of $\beta_4$ (equivalently $\gamma_4$) depends on the full model structure, not just the linear subproblem. This confirms that the main inferential advantage of the orthogonal parametrization is conditional on $\lambda$ and should be interpreted accordingly.

\subsection{Numerical results}

Table~\ref{tab:profiles} reports the conditional Fisher standard errors $\sqrt{[\sigma^2(\Phi^\top\Phi)^{-1}]{jj}}$ and $\sqrt{[\sigma^2 I_4]{jj}}$ across four conditioning regimes. In the conditional linear Gaussian setting these coincide exactly with the $\chi^2_1!=!1$ widths of the profile likelihoods shown in Fig. \ref{fig:profiles}.
\begin{table}[bt]
\centering
\small
\setlength{\tabcolsep}{4pt}
\caption{Parameter uncertainty and correlation on synthetic data: standard ($\beta$) vs.\ discrete-QR orthogonal ($\gamma$) parametrization. $\sigma = 5 \times 10^{-5}$ (0.5\,bp), maturity grid 1M, 2M, 3M, 6M, 1Y, 2Y, 3Y, 5Y, 7Y, 10Y, 20Y, 30Y. All values are analytical, computed from $\mathrm{Cov}(\hat{\beta}\mid\lambda) = \sigma^2(\Phi^\top\Phi)^{-1}$ (conditional Fisher covariance) and, for the discrete QR basis, $\mathrm{Cov}(\hat{\gamma}\mid\lambda) = \sigma^2 I_4$. True parameters: $\beta = (0.04, -0.02, 0.015, 0.008)$, $\lambda_1 = 0.6$.}
\label{tab:profiles}
\begin{tabular}{@{}lcccccccc@{}}
\toprule
Case & $\kappa(\Phi)$ & $|R_{44}|$ & $\text{std}(\hat{\beta}_1)$ & $\text{std}(\hat{\beta}_2)$ & $\text{std}(\hat{\beta}_3)$ & $\text{std}(\hat{\beta}_4)$ & $\text{std}(\hat{\gamma}_j)$ & $\max|\text{corr}(\beta)|$ \\
\midrule
\shortstack[l]{Well-separated\\($|\Delta\lambda|=0.4$)} & 39 & 0.105 & $1.0\times10^{-4}$ & $1.2\times10^{-4}$ & $2.3\times10^{-4}$ & $4.4\times10^{-4}$ & $4.8\times10^{-5}$ & 0.964 \\
\shortstack[l]{Moderate\\($|\Delta\lambda|=0.2$)} & 75 & 0.066 & $6.2\times10^{-5}$ & $7.7\times10^{-5}$ & $6.1\times10^{-4}$ & $6.7\times10^{-4}$ & $4.8\times10^{-5}$ & 0.968 \\
\shortstack[l]{Near-degenerate\\($|\Delta\lambda|=0.05$)} & 339 & 0.015 & $5.5\times10^{-5}$ & $7.1\times10^{-5}$ & $2.8\times10^{-3}$ & $2.9\times10^{-3}$ & $4.8\times10^{-5}$ & 0.999 \\
\shortstack[l]{Very degenerate\\($|\Delta\lambda|=0.01$)} & 1751 & 0.003 & $5.4\times10^{-5}$ & $7.0\times10^{-5}$ & $1.5\times10^{-2}$ & $1.5\times10^{-2}$ & $4.8\times10^{-5}$ & 1.000 \\
\bottomrule
\end{tabular}
\end{table}

The level and slope uncertainties $\text{std}(\hat{\beta}_1)$ and $\text{std}(\hat{\beta}_2)$ are essentially unchanged across regimes, while the curvature and second-hump uncertainties $\text{std}(\hat{\beta}_3)$ and $\text{std}(\hat{\beta}_4)$ increase by factors of ${\sim}65$ and ${\sim}34$, respectively, from the well-separated to the very degenerate case. Meanwhile, for the discrete QR parametrization, the uncertainty of every $\hat{\gamma}_j$ remains constant at $4.8\times10^{-5}$. Parameter correlations in the $\beta$-basis reach 0.9999 in the degenerate case; in the discrete QR $\gamma$-basis they are identically zero.

\section{\texorpdfstring{The $R_{44}$ Identifiability Diagnostic}{The R44 Identifiability Diagnostic}}\label{sec:diagnostic}

The QR factor $R$ connects the two parametrizations via $\gamma = R\beta$. The diagonal element $R_{44}$ measures the norm of the component of $\phi_4$ that is orthogonal to $\text{span}\{\phi_1, \phi_2, \phi_3\}$. As $\lambda_2 \to \lambda_1$:
\begin{equation}
|R_{44}| \to 0,
\end{equation}
because $\phi_4(\lambda_2; \tau) \to \phi_3(\lambda_1; \tau)$, making $\phi_4$ linearly dependent on the first three basis functions.

The same interpretation applies to the finite-horizon analytical factor $R_T = L_T^\top$ from Proposition~\ref{prop:finiteT}: the entry $(R_T)_{44}$ measures the orthogonal distance of $\phi_4$ from $\mathrm{span}\{\phi_1,\phi_2,\phi_3\}$ in $L^2(0,T)$. By Remark~\ref{rem:finitebasis}, this diagnostic is horizon-dependent. By Remark~\ref{rem:infinitebasis}, its infinite-horizon limit is canonical only after passing to the decaying $3\times3$ subspace, since the full $4\times4$ basis does not survive on $(0,\infty)$.

\begin{figure}[tb]
\centering
\includegraphics[width=0.65\textwidth]{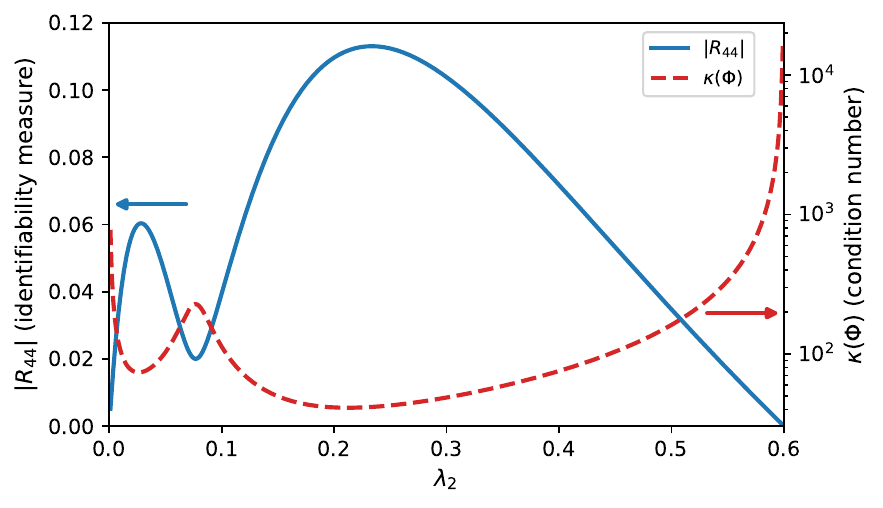}
\caption{Identifiability diagnostic $|R_{44}|$ and condition number $\kappa(\Phi)$ as functions of $\lambda_2$ for fixed $\lambda_1 = 0.6$. As $\lambda_2 \to \lambda_1$, $|R_{44}| \to 0$ and $\kappa \to \infty$. For $\lambda_2 \to 0$ the condition number also diverges.}
\label{fig:r44}
\end{figure}

\paragraph{Quantitative threshold}
Since $\beta = R^{-1}\gamma$ and $\mathrm{Cov}(\hat{\gamma}\mid\lambda) = \sigma^2 I_4$, the conditional uncertainty of $\hat{\beta}_4$ satisfies
\begin{equation}\label{eq:r44_threshold}
\mathrm{std}(\hat{\beta}_4 \mid \lambda) = \sigma \|e_4^\top R^{-1}\| \geq \frac{\sigma}{|R_{44}|},
\end{equation}
where $e_4$ is the fourth standard basis vector and the inequality follows from $|[R^{-1}]_{44}| = 1/|R_{44}|$ for upper-triangular $R$. For a desired maximum parameter uncertainty $\delta$ (e.g., $\delta = 10\sigma$, corresponding to a signal-to-noise ratio of~10), model reduction from Svensson to Nelson--Siegel is warranted when
\begin{equation}
|R_{44}| < \frac{\sigma}{\delta}.
\end{equation}
With $\sigma = 5\times10^{-5}$ (0.5\,bp) and $\delta = 5\times10^{-4}$ (5\,bp), the threshold is $|R_{44}| < 0.1$. For the four cases in Table~\ref{tab:profiles}, $|R_{44}|$ takes values $0.105, 0.066, 0.015, 0.003$. The well-separated case ($|R_{44}| = 0.105$) barely exceeds the threshold, while all three closer-spaced cases fall below it. This illustrates the sensitivity of the diagnostic: even moderate proximity ($|\Delta\lambda| = 0.2$) produces $|R_{44}|$ below the threshold, indicating that the Svensson extension is informative only when the decay parameters are well separated.

Accordingly, $|R_{44}| \geq \sigma/\delta$ indicates that the Svensson extension remains informative and the full four-component model is justified, whereas $|R_{44}| < \sigma/\delta$ indicates that the fourth component is effectively redundant and reduction to Nelson--Siegel is warranted. Concretely, with $\sigma = 0.5$\,bp and desired precision $\delta = 5$\,bp, reduction is indicated when $|R_{44}| < 0.1$.

Figure~\ref{fig:r44} shows $|R_{44}|$ and $\kappa(\Phi)$ as functions of $\lambda_2$ for fixed $\lambda_1 = 0.6$. The condition number diverges as $|R_{44}| \to 0$, confirming the diagnostic.

\begin{figure}[tb]
\centering
\includegraphics[width=\textwidth]{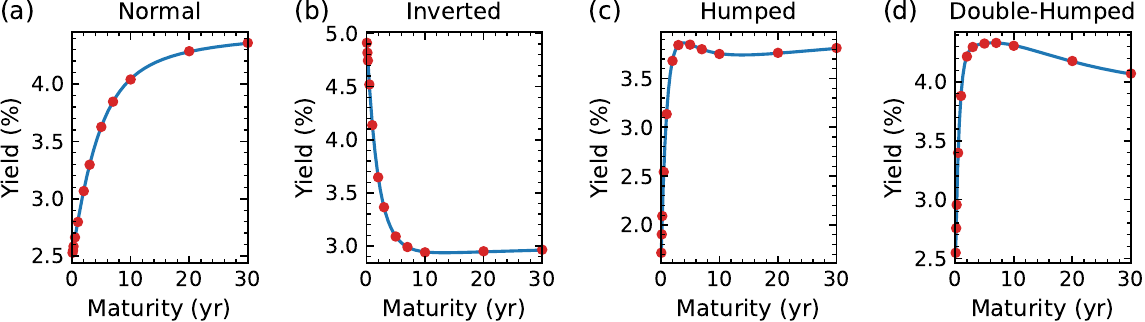}
\caption{Example yield curves from four synthetic regimes: (a) normal upward sloping, (b) inverted with negative initial slope, (c) single-humped and (d) double humped with positive initial slope.}
\label{fig:curves}
\end{figure}

\section{Regularization Analysis: Standard vs.\ Orthogonal Bases}\label{sec:regularization}

\subsection{Orthogonal ridge regression}

For fixed~$\lambda$, in the orthogonal basis, penalizing $\|\gamma\|^2$ yields
\begin{equation}
\hat{\gamma}_\alpha = \arg\min_\gamma \|\Psi(\lambda)\gamma - \hat{y}\|^2 + \alpha\|\gamma\|^2 = \frac{1}{1+\alpha}\Psi(\lambda)^\top\hat{y}.
\end{equation}
This shrinks each component by the same factor $(1+\alpha)^{-1}$, providing isotropic (uniform) regularization in the decorrelated space. The corresponding penalty in the original basis is
\begin{equation}
\|\gamma\|^2
= \|R(\lambda)\beta\|^2
= \beta^\top R(\lambda)^\top R(\lambda)\,\beta
= \beta^\top(\Phi(\lambda)^\top\Phi(\lambda))\beta,
\end{equation}
which penalizes directions proportionally to how well-determined they already are.

\subsection{Standard ridge regression}

In contrast, standard ridge regression penalizes $\|\beta\|^2$:
\begin{equation}
\hat{\beta}_\alpha = \arg\min_\beta \|\Phi\beta - \hat{y}\|^2 + \alpha\|\beta\|^2 = (\Phi^\top\Phi + \alpha I)^{-1}\Phi^\top\hat{y}.
\end{equation}
Via the SVD of $\Phi = U\Sigma V^\top$, this applies filter factors $f_i = \sigma_i^2/(\sigma_i^2 + \alpha)$ that differentially suppress ill-conditioned directions: small singular values receive stronger shrinkage. Standard ridge therefore applies stronger shrinkage to ill-conditioned directions via the SVD filter factors, targeting the noise-amplifying directions more effectively~\citep{hoerl1970}.

\subsection{Regularization choices \& connection to regularized variable projection}
The choice between penalizing $\|\beta\|^2$ and $\|\gamma\|^2 = \|R(\lambda)\beta\|^2$ is a specific instance of the general-form Tikhonov framework $\min_\beta \|\Phi(\lambda)\beta - \hat{y}\|^2 + \alpha\|L\beta\|^2$ with regularization matrix $L = I$ (standard ridge) vs.\ $L = R(\lambda)$ (orthogonal ridge, at the current value of~$\lambda$). Thus uniform shrinkage in $\gamma$ becomes a $\lambda$-dependent general-form penalty when written in the original $\beta$ coordinates. Recent work on regularized variable projection~\citep{espanol2023} develops VARPRO with general-form Tikhonov regularization on the linear parameters, showing that the choice of~$L$ fundamentally changes the reduced problem's landscape. Comerso~Salzer et~al.~\citep{comersosalzer2026} extend this framework to simultaneous regularization of both linear and nonlinear parameters, while Espa\~nol and Jeronimo~\citep{chung2025} provide convergence guarantees for the Gauss--Newton method on the regularized reduced problem. Our analysis provides a concrete instance of these general results: the NSS model demonstrates that $L = I$ (standard ridge) outperforms $L = R(\lambda)$ (orthogonal ridge) precisely because the ill-conditioned directions of $\Phi(\lambda)$ align with the directions that need the most shrinkage. This is complementary to the SVD-based sparse modeling approach of \citet{guterding2023}: there, singular vectors and $L^1$ regularization are used to stabilize an ill-posed inverse problem for option-implied densities, whereas here the objective is to retain the NSS model exactly while exposing its near-redundant directions and identifiability structure.

\begin{figure}[tb]
\centering
\includegraphics[width=0.85\textwidth]{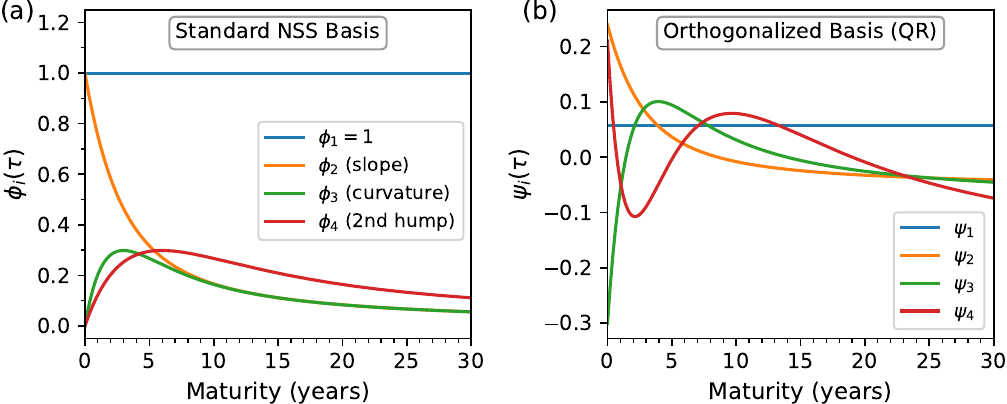}
\caption{Basis functions as a function of maturity for (a) the standard NSS model and (b) the discrete-QR orthogonalized NSS model.}
\label{fig:basis}
\end{figure}

As our experiments confirm (see Section~\ref{sec:experiments}), standard ridge outperforms orthogonal ridge in the degenerate regime because it targets the noise-amplifying directions more effectively. The orthogonal ridge penalty $\beta^\top(\Phi^\top\Phi)\beta$ upweights directions that are already well-determined while leaving the ill-conditioned directions relatively unpenalized---precisely the opposite of what is needed.

The SVD analysis and numerical results favor standard ridge regression $\|\beta\|^2$ for stabilizing the inner linear subproblem when regularization is needed, with the regularization parameter $\alpha$ chosen by generalized cross-validation~\citep{golub1979}; recent work~\citep{bellec2025} shows that standard GCV can be biased in finite samples with correlated features. The orthogonal basis provides a complementary diagnostic: when $|R_{44}|$ is small, truncation to $k=3$ components (equivalent to model reduction from NSS to NS) is an effective alternative to continuous regularization. This connects to the truncated SVD approach for discrete ill-posed problems~\citep{hansen1992}.

In practice, yield curve fitting almost always uses weighted least squares (WLS) with weight matrix $W = \mathrm{diag}(w_1, \ldots, w_m)$ to account for heteroscedastic pricing errors across maturities~\citep{bis2005}. The orthogonal reparametrization can be extended directly: QR decomposition of $W^{1/2}\Phi(\lambda)$ produces an orthonormal basis $\Psi_W$ for the weighted problem, with $\mathrm{Cov}(\hat{\gamma}_W \mid \lambda) = \sigma^2 I_4$ under the weighted model $\varepsilon \sim \mathcal{N}(0, \sigma^2 W^{-1})$. The $R_{44}$ diagnostic carries over unchanged: it still measures the near-redundancy of the fourth basis function, now in the weighted inner product. A real-data WLS application using duration-inverse weights following ECB practice is left for future work.

\begin{figure}[tb]
\centering
\includegraphics[width=0.7\textwidth]{fig4_condition_landscape.png}
\caption{Condition number landscape $\log_{10}\kappa(\Phi(\lambda_1, \lambda_2))$ for the standard NSS basis on the finite plotting grid. The exact diagonal $\lambda_1 = \lambda_2$ and boundary singularities are excluded from the plotted finite values; analytically, $\kappa(\Phi(\lambda_1,\lambda_2))$ diverges as these singular sets are approached.}
\label{fig:cond}
\end{figure}

\section{Numerical Experiments}\label{sec:experiments}

This section is organized into controlled computational evidence and application evidence. The synthetic experiments test the conditioning, identifiability, and regularization claims in regimes with known ground truth, while the Treasury study provides a nontrivial long-history calibration example with time-varying nonlinear parameters and an imperfect maturity grid.

\subsection{Synthetic data generation}

We test on synthetic yield curves generated from the NSS model with known parameters across eight regimes (normal, flat, inverted, humped, double-humped, near-degenerate, small-$\lambda$, large-$\lambda$). Maturities correspond to US Treasury standard maturities: 1M, 2M, 3M, 6M, 1Y, 2Y, 3Y, 5Y, 7Y, 10Y, 20Y, 30Y. Gaussian noise with $\sigma = 5\times10^{-5}$ (0.5 basis points) is added, consistent with the precision of high-quality sovereign bond yield fits where benchmark U.S.\ Treasury bid--ask spreads are around two basis points~\citep{campbell2014}.

The choice $\sigma = 5\times10^{-5}$ is intended as an optimistic high-precision benchmark for liquid Treasury markets. It is smaller than the roughly two-basis-point bid--ask spreads cited above, but is motivated by the fact that benchmark Treasury yield-curve smoothers can fit the cross section closely; G\"urkaynak--Sack--Wright report that their six-parameter Treasury curve closely fits the cross section and that average absolute yield prediction errors are generally small over the sample~\citep{gurkaynak2007}. 
These conclusions are noise-level-independent by construction. The identity $\mathrm{std}(\hat\gamma_j \mid \lambda) = \sigma$ and the bound                 
  $\mathrm{std}(\hat\beta_4 \mid \lambda)/\mathrm{std}(\hat\gamma_4 \mid \lambda) \geq 1/|R_{44}|$ hold exactly, with the latter depending only on $\lambda$. With $\sigma = 5$ bp and $\delta = 5$ bp, the threshold value is $\sigma/\delta = 1$, so reduction is indicated whenever $|R_{44}| < 1$. Because $|R_{44}| \leq \|\phi_4\|_2 \leq \sqrt{m}/e$ (since $\phi_4(\tau)$ peaks pointwise at $1/e$), with $\sqrt{m}/e \approx 1.27$ for the US Treasury grid, $|R_{44}|$ is bounded by a modest constant; empirically (Table~\ref{tab:profiles}) all four cases fall below 0.11, well under 1; the Svensson extension is rarely informative for this signal-to-noise ratio.
\subsection{Basis function comparison}

Figure~\ref{fig:basis} shows the standard NSS basis functions and their discrete-QR orthogonalization for $\lambda_1 = 0.6$, $\lambda_2 = 0.3$. The standard basis exhibits visible similarity between $\phi_2$ and $\phi_3$; the orthogonalized basis produces visually distinct functions.

\subsection{Condition number landscape}

Figure~\ref{fig:cond} shows $\log_{10}\kappa(\Phi)$ for the standard NSS basis over the $(\lambda_1, \lambda_2)$ parameter space. The finite plotting grid shows large but finite values near the diagonal $\lambda_1 = \lambda_2$ and at the boundaries. Analytically, the condition number diverges on these singular sets and can exceed $10^{15}$ when evaluated sufficiently close to them. The discrete QR basis has $\kappa(\Psi) = 1$ everywhere by construction.

\subsection{Correlation matrices}

\begin{figure}[tb]
\centering
\includegraphics[width=\textwidth]{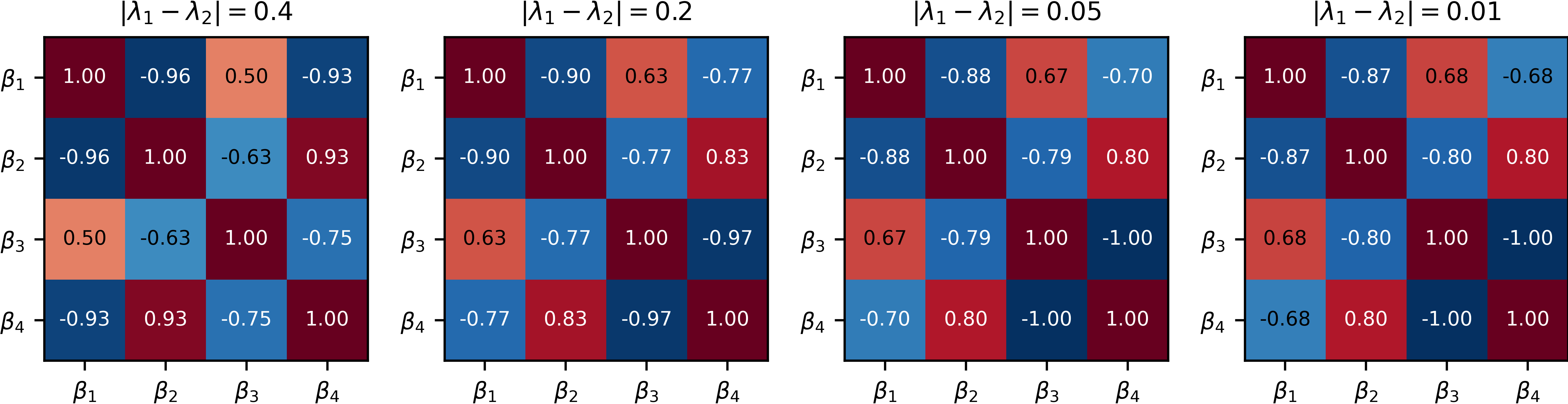}
\caption{Correlation matrices for the standard $\beta$ parametrization across four conditioning regimes. The maximum absolute off-diagonal correlation reported in Table~\ref{tab:profiles} is computed from these matrices. In the orthogonal $\gamma$ parametrization the correlation matrix is diagonal by construction.}
\label{fig:corr}
\end{figure}

Figure~\ref{fig:corr} displays the parameter correlation matrices for four conditioning regimes. In the standard parametrization, correlations exceed 0.99 in the near-degenerate case. The orthogonal parametrization yields zero off-diagonal correlations by construction.

\subsection{Profile likelihoods}

\begin{figure}[tb]
\centering
\includegraphics[width=\textwidth]{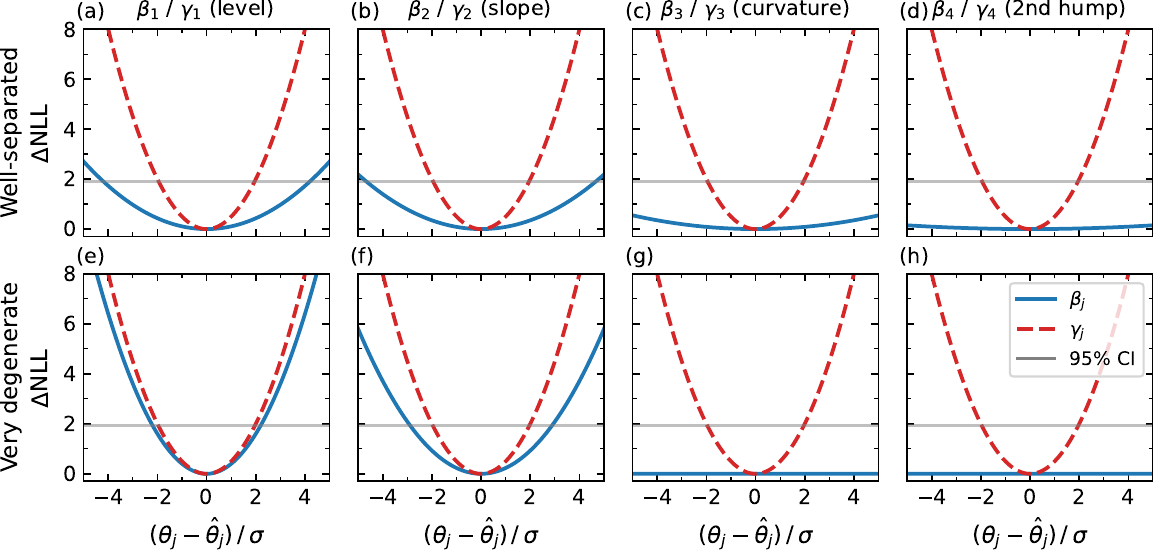}
\caption{Profile negative log-likelihood $\Delta\text{NLL}$ for the four linear parameters: level ($\beta_1/\gamma_1$), slope ($\beta_2/\gamma_2$), curvature ($\beta_3/\gamma_3$), and second hump ($\beta_4/\gamma_4$). Blue solid: standard $\beta_j$; red dashed: orthogonal $\gamma_j$. (a-d) well-separated case ($\lambda_1 = 0.6$, $\lambda_2 = 0.2$). (e-h) very degenerate case ($\lambda_1 = 0.6$, $\lambda_2 = 0.59$). The $x$-axis shows $(\theta_j - \hat{\theta}_j)/\sigma$ so both parametrizations share the same scale; the grey horizontal line marks the 95\% confidence threshold ($\frac{1}{2}\chi^2_1 = 1.92$). In the degenerate case, $\beta_3$ and $\beta_4$ profiles are flat (non-identifiable) while all $\gamma$ profiles retain full curvature. Profiles are conditional on fixed~$\lambda$.}
\label{fig:profiles}
\end{figure}

Figure~\ref{fig:profiles} shows the profile negative log-likelihood Eq. (\ref{eq:NLL}) for each of the four linear parameters.
This quantity equals $\frac{1}{2}\Delta\chi^2$ and attains its minimum (zero) at the maximum likelihood estimator (MLE, which is the least squares estimate); a parabolic profile indicates a well-identified parameter, while a flat profile signals non-identifiability. Each parameter's deviation from its MLE is normalized by the noise standard deviation~$\sigma$, placing both the standard ($\beta_j$, blue solid) and orthogonal ($\gamma_j$, red dashed) profiles on the same horizontal scale. These profiles are conditional on fixed~$\lambda$: the upper row uses $\lambda_1=0.6,\lambda_2=0.2$, while the lower row uses $\lambda_1=0.6,\lambda_2=0.59$. The upper row shows the well-separated case ($|\lambda_1 - \lambda_2| = 0.4$): all eight profiles are parabolic, though the $\gamma$ profiles are steeper because $\text{std}(\hat{\gamma}_j) = \sigma$ while $\text{std}(\hat{\beta}_j) > \sigma$ due to multicollinearity. The lower row shows the very degenerate case ($|\lambda_1 - \lambda_2| = 0.01$): the $\beta_3$ and $\beta_4$ profiles become completely flat---$\Delta\text{NLL}$ does not increase even at $\pm 5\sigma$ from the MLE---while all four $\gamma$ profiles retain full curvature, identical to the well-separated case.

\begin{figure}[tb]
\centering
\includegraphics[width=\textwidth]{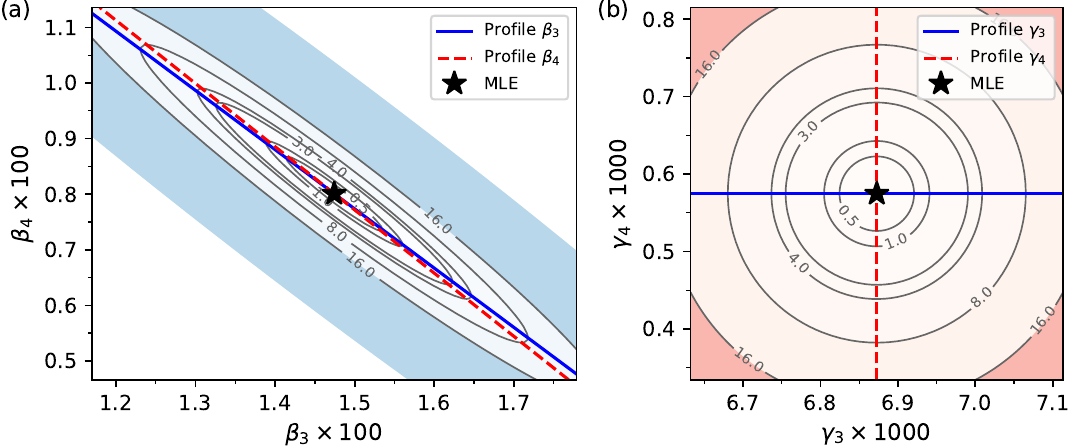}
\caption{Two-dimensional $\Delta\text{NLL}$ landscape over parameter pairs $(\beta_3, \beta_4)$ (a) and $(\gamma_3, \gamma_4)$ (b) in the well-separated regime ($\lambda_1 = 0.6$, $\lambda_2 = 0.4$). Contour labels show $\Delta\text{NLL}$ values; the $3.0$ contour corresponds to the joint $95\%$ confidence region ($\tfrac{1}{2}\chi^2_{2,0.95} \approx 3.0$). Profile paths (solid: parameter~3 profiled; dashed: parameter~4 profiled) show the optimizer's trajectory when fixing one parameter and optimizing the rest. Stars mark the MLE.}
\label{fig:landscape}
\end{figure}

Figure~\ref{fig:landscape} provides a complementary two-dimensional view: the profiled $\Delta\text{NLL}$ landscape over the $(\beta_3, \beta_4)$ and $(\gamma_3, \gamma_4)$ planes in the well-separated regime ($\lambda_1 = 0.6$, $\lambda_2 = 0.4$). In the standard basis, the contours form elongated ellipses along the ridge $\beta_3 + \beta_4 \approx \text{const}$, and the profile paths for $\beta_3$ and $\beta_4$ nearly coincide as they trace the valley floor. This problem is strongly amplified when entering the near-degenerate regime. In the orthogonal basis, contours are circular and profile paths are axis-aligned---each parameter is independently identifiable.

\subsection{Long-history Treasury calibration study}\label{sec:timeseries}

To test whether the orthogonal parameter series remain useful when the decay parameters vary from day to day, we fit the NSS model independently to the daily US Treasury constant-maturity history contained in the repository data set. A data limitation is that the 20Y tenor is unavailable for a long interval (1987-01-01 to 1993-09-30). To avoid an artificial break in the complete-case sample, we define a fixed reduced 9-tenor grid
\[
3\text{M},\, 6\text{M},\, 1\text{Y},\, 2\text{Y},\, 3\text{Y},\, 5\text{Y},\, 7\text{Y},\, 10\text{Y},\, 30\text{Y},
\]
which yields 11{,}633 daily curves from 1981-09-01 to 2026-04-02 (note that a year has $\sim250$ trading days). As a robustness check, we also consider a complete 11-tenor sample from 2001-07-31 onward. \ref{app:tenor-robustness} adds a complementary post-gap comparison on the common 1993-10-01 to 2026-04-02 window, where we contrast this reduced grid with the maximal fixed 10-tenor grid that reinstates the 20Y maturity after its return. Each day is fitted independently: the first day uses a global search over~$\lambda$, and subsequent days use the previous day's estimate as a warm start. Since $\Phi(\lambda)=\Psi(\lambda)R(\lambda)$, the reduced objective
\[
H_t(\lambda)=\min_{\beta}\|\Phi(\lambda)\beta-y_t\|_2^2=\min_{\gamma}\|\Psi(\lambda)\gamma-y_t\|_2^2
\]
is identical in the $\beta$- and $\gamma$-parametrizations, so the daily sequence $\lambda_t$ is unchanged by orthogonalization; only the reported linear parameters differ.
Whenever the classical parameters are needed for comparison, they are recovered from the same daily fit by the upper-triangular solve $R(\lambda_t)\beta_t=\gamma_t$, i.e.\ $\beta_t = R(\lambda_t)^{-1}\gamma_t$, exactly as in Section~\ref{sec:fit-gamma}. This back-transformation is only numerically reliable when $|R_{44}|$ is not too small; when the fourth direction is weakly identified, the recovered $\beta_t$ values---especially $\beta_{3,t}$ and $\beta_{4,t}$---should be treated as reporting transforms rather than stable state variables.

For a parameter time series $p_t$, we summarize smoothness by
\[
\rho(p)=\frac{\mathrm{std}(\Delta p_t)}{\mathrm{std}(p_t)},
\qquad
J(p)=\frac{q_{0.95}(|\Delta p_t|)}{\mathrm{IQR}(p_t)},
\]
where lower values indicate smoother evolution relative to the scale of the series. To quantify time comparability of the moving orthogonal basis, we define the basis-rotation diagnostic
\[
B_t=\|\Psi(\lambda_{t-1})^\top\Psi(\lambda_t)-I\|_F.
\]

\begin{figure}[tbp]
\centering
\includegraphics[width=\textwidth]{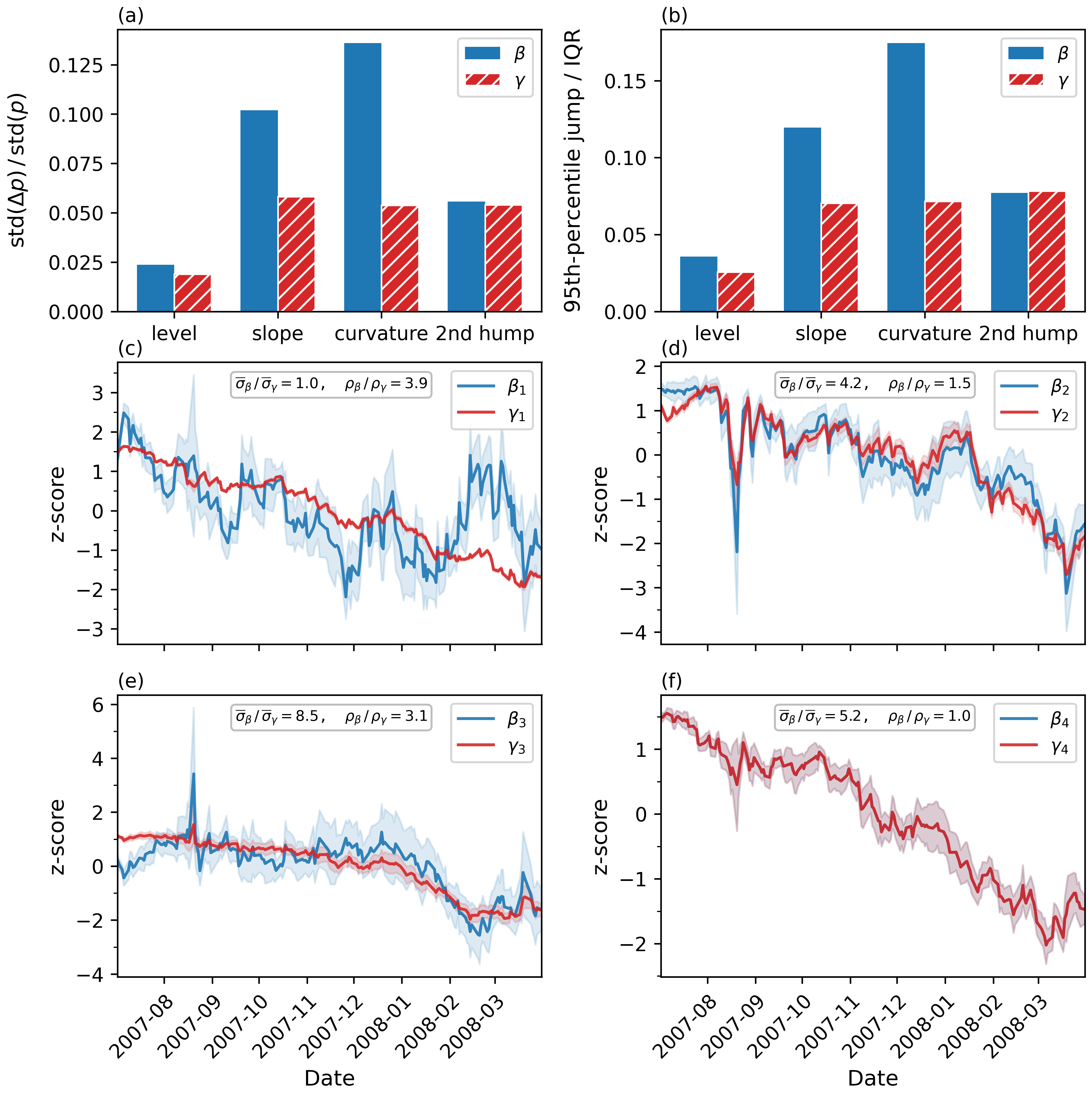}
\caption{Daily US Treasury application on the fixed reduced 9-tenor grid. (a, b) aggregate stability diagnostics for the classical parameters $\beta_t$ and orthogonal parameters $\gamma_t$ (1981--2026). (c-f) standardized crisis-window overlays for level, slope, curvature, and second hump during 2007--2008, with $\pm 3\sigma$ Fisher-information uncertainty bands (shaded) and per-panel ratios $\bar{\sigma}_\beta/\bar{\sigma}_\gamma$ and $\rho_\beta/\rho_\gamma$. The orthogonal parameters are visibly smoother and carry tighter per-day confidence bounds, particularly for the curvature and second-hump components.}
\label{fig:daily-stability}
\end{figure}

On the primary reduced 9-tenor sample, the orthogonal parameter series are smoother than the classical parameter series for the first three components: the roughness ratios $\rho(\beta_j)/\rho(\gamma_j)$ are $1.25$, $1.75$, $2.53$, and $1.03$ for $j=1,\dots,4$, while the jump ratios $J(\beta_j)/J(\gamma_j)$ are $1.38$, $1.69$, $2.44$, and $0.99$. Thus the gain is largest in the third component, still material in the first two, and essentially neutral for the fourth. On the 11-tenor robustness sample the same qualitative conclusion holds, with roughness improvements of $4.06$, $2.25$, $1.57$, and $1.05$. Figure~\ref{fig:daily-stability} shows both the aggregate stability diagnostics and the crisis-window overlays for 2007--2008: across components 1--3, the $\gamma_t$ series moves more smoothly than the corresponding $\beta_t$ series while preserving the same fitted curves.

\begin{figure}[tbp]
\centering
\includegraphics[width=\textwidth]{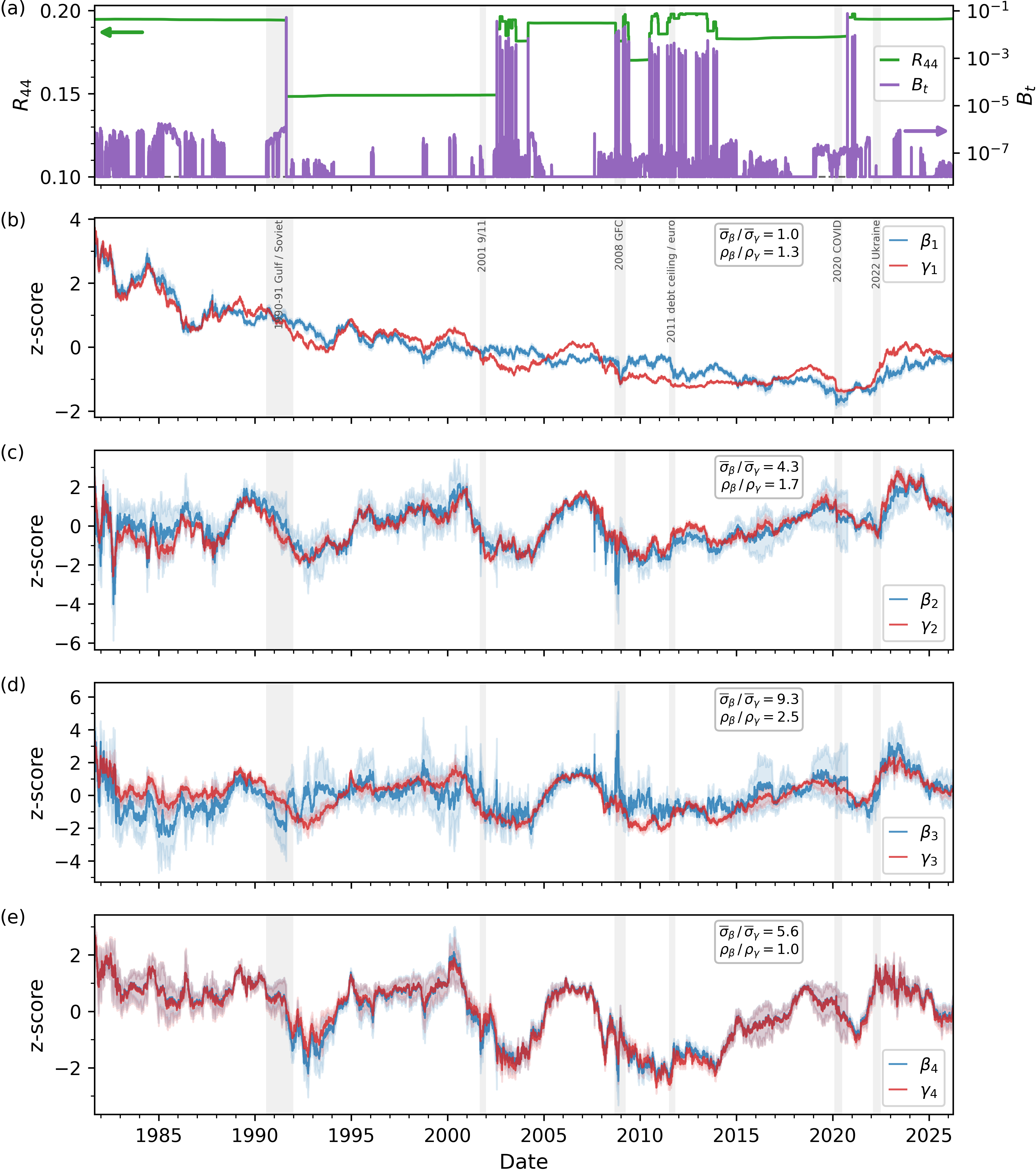}
\caption{Daily identifiability and long-run state representation on the fixed reduced 9-tenor grid (1981--2026). (a) the identifiability diagnostic $R_{44}$ (left axis) and basis-rotation diagnostic $B_t=\|\Psi(\lambda_{t-1})^\top\Psi(\lambda_t)-I\|_F$ (right axis, log scale). (b-e) standardized full-sample series for level, slope, curvature, and second hump with $\pm 3\sigma$ Fisher-information uncertainty bands (shaded). Each panel annotates the raw Fisher standard error ratio $\bar{\sigma}_\beta/\bar{\sigma}_\gamma$ and the roughness ratio $\rho_\beta/\rho_\gamma$. Ratios above $1$ would indicate worse uncertainty or roughness in the classical basis. Visible band widths in z-score space also reflect the different temporal scales of $\beta$ and $\gamma$.}
\label{fig:daily-history-summary}
\end{figure}

Interpretability requires a separate check because the orthogonal basis depends on $\lambda_t$, so the same-index $\gamma_j$ values are meaningful over time only if the basis itself moves slowly. On the primary reduced 9-tenor sample this concern is empirically mild. The 95th percentile of $B_t$ is $5.88\times 10^{-7}$, the 99th percentile is $1.32\times 10^{-6}$, and the maximum is only $0.0775$; no day exceeds $0.1$. Figure~\ref{fig:daily-history-summary} combines this diagnostic with the full-history component trajectories. The reduced 9-tenor sample also stays away from the severe identifiability breakdown of the complete-case 10-tenor sample: $R_{44}$ never falls below $0.148$. On the 11-tenor robustness sample, rare large-rotation dates do occur (the maximum is $0.62$, concentrated around the 2007--2009 crisis), so large values of $B_t$ should still be flagged in operational use. On the long reduced-grid sample, however, the orthogonal parameter series remain usable as a daily state representation rather than merely as within-day diagnostics.

The long-horizon standardized trajectories in Figure~\ref{fig:daily-history-summary} reveal an additional benefit: the orthogonal parameter series make low-frequency shifts easier to read. The same broad transitions are visible in both parametrizations, but the orthogonal parameter series exhibit less short-run oscillation and less cross-contamination across all four components.

Figure~\ref{fig:daily-history-summary} also overlays $\pm 3\sigma$ Fisher-information uncertainty bands, computed per day from $\mathrm{std}(\hat{\beta}_j\mid\lambda_t) = \sigma_t\sqrt{[(\Phi_t^\top\Phi_t)^{-1}]_{jj}}$ and $\mathrm{std}(\hat{\gamma}_j\mid\lambda_t) = \sigma_t$, where $\sigma_t^2 = \mathrm{RSS}_t/(m-4)$. The annotated ratios $\bar{\sigma}_\beta/\bar{\sigma}_\gamma$ report the raw (unscaled) Fisher standard error ratio, which isolates the statistical uncertainty effect: $1.0\times$ for level, $4.3\times$ for slope, $9.3\times$ for curvature, and $5.6\times$ for the second hump. The roughness ratios $\rho_\beta/\rho_\gamma$ are $1.3\times$, $1.7\times$, $2.5\times$, and $1.0\times$. Curvature is the worst case on both measures, consistent with $\beta_3$ absorbing the largest share of the near-degeneracy when $\lambda_1\approx\lambda_2$. The second hump shows a notable asymmetry: its per-day uncertainty is $5.6\times$ worse in the classical basis, yet its roughness ratio is only $1.0\times$---the extra uncertainty does not manifest as rougher dynamics but rather as less trustworthy point estimates. Note that the visible band widths in the z-score panels also reflect the different temporal scales of $\beta$ and $\gamma$ (since $\gamma = R\beta$, the orthogonal parameters live on a different scale); the annotated raw ratios should therefore be read as the statistical quantity of interest rather than as a visual legend for band width.

Since the fitted yield curve $\hat{y}_t = \Phi_t\hat{\beta}_t = \Psi_t\hat{\gamma}_t$ is identical in both parametrizations, the fit quality is unchanged by orthogonalization. Table~\ref{tab:fit-quality} confirms this on the primary 9-tenor sample. The overall median RMSE is 4.66\,bp, and the 95th percentile is 10.39\,bp. These values are of the same order of magnitude as reported yield-curve fitting errors in practice, especially at longer maturities. Splitting by the median of $R_{44}$ reveals no meaningful degradation in either regime: the median RMSE is $4.41$\,bp in well-conditioned periods and $4.87$\,bp in ill-conditioned periods, and the rank-order correlation between daily RMSE and $R_{44}$ is only $-0.076$. The per-maturity breakdown shows that the largest RMSE contributions come from the 30Y ($8.26$\,bp) and 10Y ($7.10$\,bp) maturities, reflecting the well-known difficulty of capturing the long end with only four factors. The improved parameter uncertainty from orthogonalization is therefore obtained at no cost to fit quality.

\begin{table}[tb]
\centering
\small
\setlength{\tabcolsep}{5pt}
\caption{NSS fit quality on the US Treasury 9-tenor grid (1981--2026). Top panel: aggregate fit statistics split by the median of $R_{44}$. Bottom panel: per-maturity mean bias and RMSE. All values in basis points. The correlation between daily RMSE and $R_{44}$ is $-0.076$, confirming that fit quality is independent of conditioning.}
\label{tab:fit-quality}
\begin{tabular}{@{}lccc@{}}
\toprule
& All days & $R_{44} >$ median & $R_{44} \leq$ median \\
& ($N = 11{,}145$) & ($N = 5{,}572$) & ($N = 5{,}573$) \\
\midrule
Median RMSE (bp) & 4.66 & 4.41 & 4.87 \\
95th pctl RMSE (bp) & 10.39 & 10.85 & 10.16 \\
Max RMSE (bp) & 17.72 & 17.72 & 14.37 \\
Median max$|e_i|$ (bp) & 8.41 & 7.48 & 9.24 \\
\bottomrule
\end{tabular}

\vspace{6pt}

\begin{tabular}{@{}lrrrrrrrrr@{}}
\toprule
Maturity & 3M & 6M & 1Y & 2Y & 3Y & 5Y & 7Y & 10Y & 30Y \\
\midrule
Mean bias (bp) & $-0.82$ & $+2.49$ & $-3.21$ & $+2.36$ & $+0.58$ & $-1.68$ & $+0.46$ & $-2.75$ & $+2.58$ \\
RMSE (bp) & 2.42 & 6.48 & 5.03 & 5.96 & 4.07 & 4.94 & 6.77 & 7.10 & 8.26 \\
\bottomrule
\end{tabular}
\end{table}

To quantify this low-frequency structure, we also solve a multivariate piecewise-constant mean problem on the monthly 4-dimensional $\beta$ and $\gamma$ histories by dynamic programming, which is globally optimal conditional on the number of changepoints. On this sample, a plain Bayesian information criterion (BIC) is too permissive and keeps favoring additional breaks up to the truncation boundary, so we use the elbow of the exact log-SSE path as a practical complexity choice. This yields six changepoints for $\beta$ and seven for $\gamma$. For $\beta$ they occur at 1991-09, 1994-12, 2001-09, 2005-02, 2008-02, and 2016-02, while for $\gamma$ they occur at 1991-05, 1994-10, 2001-09, 2005-01, 2008-01, 2015-12, and 2022-07. The first five break dates are nearly identical in both parametrizations. The important difference is therefore not the historical chronology but the sharpness of the regime representation: at the $\gamma$ seven-break complexity, the within-segment SSE is $753.6$ for $\beta$ (at six breaks) and $534.3$ for $\gamma$ (at seven breaks), a reduction of about $29.1\%$. In that sense, the orthogonal parameter series do not imply a different macro-financial history; they provide a cleaner low-frequency representation of the same history.

\begin{figure}[tbp]
\centering
\includegraphics[width=\textwidth]{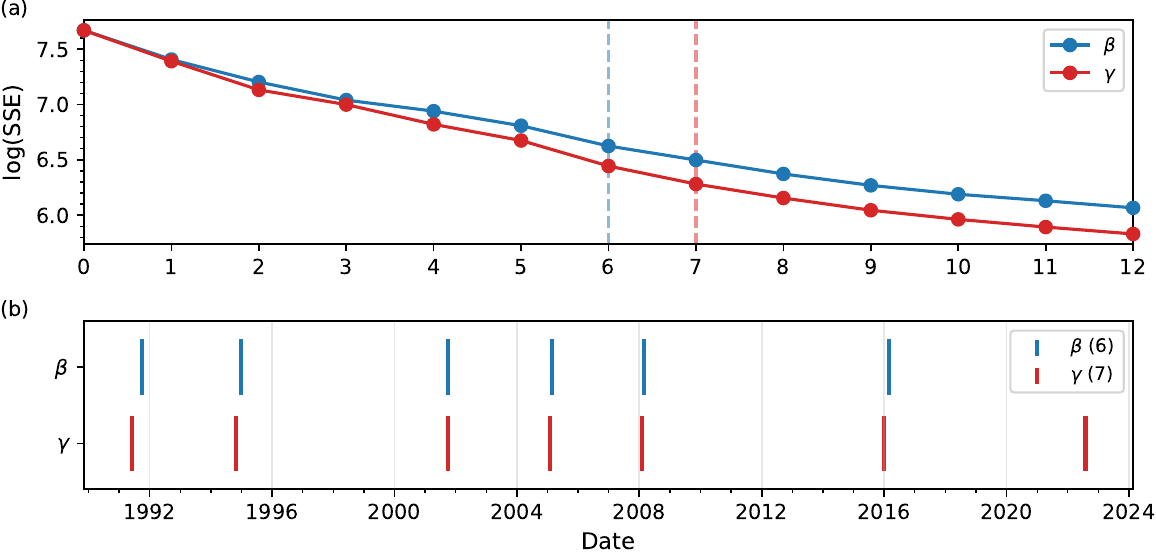}
\caption{Exact monthly changepoint analysis on the fixed reduced 9-tenor grid. (a) exact dynamic-programming cost paths for the multivariate piecewise-constant mean model, with the elbow-selected complexity marked for $\beta$ and $\gamma$. (b) the resulting elbow-selected changepoint dates. Both parametrizations imply the same broad seven-regime chronology, but the orthogonal parameter series achieve a lower within-segment SSE at that common complexity.}
\label{fig:daily-changepoints}
\end{figure}

\section{Discussion}\label{sec:discussion}

\subsection{Numerical, statistical, and predictive stability}\label{sec:stability}

Three levels should be distinguished when assessing the effect of orthogonalization:

\begin{enumerate}
\item \textbf{Algorithm accuracy (numerical stability).} The QR-based computation of the inner linear solve has forward error $O(\kappa(\Phi)\,\varepsilon_{\mathrm{mach}})$, compared to $O(\kappa(\Phi)^2\,\varepsilon_{\mathrm{mach}})$ for the normal equations approach $\hat{\beta} = (\Phi^\top\Phi)^{-1}\Phi^\top\hat{y}$~\citep{higham2002,bjorck1996}. Note that modern implementations (including our code's standard solver, which uses SVD-based LAPACK routines internally) already achieve $O(\kappa(\Phi)\,\varepsilon_{\mathrm{mach}})$ accuracy. The practical advantage of the explicit QR approach is therefore primarily versus na\"ive normal-equations implementations, which are still common in production yield curve fitting code. The QR decomposition additionally provides the $R_{44}$ diagnostic at no extra cost.

\item \textbf{Parameter uncertainty (statistical stability).} Conditional on fixed $\lambda$, the covariance $\mathrm{Cov}(\hat{\beta}\mid\lambda) = \sigma^2(\Phi^\top\Phi)^{-1}$ is an intrinsic property of the model and data, unchanged by the algorithm. Orthogonalization does not reduce uncertainty in $\hat{\beta}$; in the discrete QR $\gamma$-parametrization it merely rewrites the same conditional uncertainty as $\mathrm{Cov}(\hat{\gamma}\mid\lambda) = \sigma^2 I_4$. When $\lambda$ is estimated jointly, Proposition~\ref{prop:fullcov} shows that the orthogonal coordinates acquire the additional inflation term $\sigma^2 C S^{-1}C^\top$, and Corollary~\ref{cor:beta-cov} shows how this nonlinear uncertainty propagates further into the recovered classical parameters through the $\lambda$-dependence of~$R$.

\item \textbf{Prediction uncertainty.} The predicted yield curve $\hat{y}(\tau)$ and its prediction intervals are identical in both parametrizations, since $\Phi\hat{\beta} = \Psi\hat{\gamma}$.
\end{enumerate}

The practical benefit of orthogonalization is primarily diagnostic and computational, not statistical: it improves algorithm accuracy (point~1), provides interpretable uncertainty decomposition both conditionally (Proposition~\ref{prop:fisher}) and at first order under joint estimation (Proposition~\ref{prop:fullcov}), and enables the $R_{44}$ diagnostic (Section~\ref{sec:diagnostic}).

The historical application in Section~\ref{sec:timeseries} adds a fourth, operational layer: even when $\lambda$ is re-estimated every day, the orthogonal parameter series provide smoother reported state variables than the classical $\beta$ series, and on a fixed reduced maturity grid the basis rotation is so small that the same-index $\gamma_j$ remain comparable over time. Orthogonalization is therefore relevant not only for one-date diagnostics but also for restartable day-to-day calibration workflows. The exact monthly changepoint analysis serves as a robustness check on that conclusion: the broad regime chronology is almost unchanged by the reparametrization, but the orthogonal parameter series admit a materially cleaner piecewise-constant representation of those regimes.

It also sharpens low-frequency interpretation. The long-run plots do not suggest that orthogonalization erases genuine regime changes; rather, they show that broad shifts become easier to isolate when the short-run noise of the linear parameters is reduced. For monitoring and surveillance applications, this cleaner separation may be more useful than the classical economic labels attached to $\beta_1,\ldots,\beta_4$.

\paragraph{Relationship to PCA-based yield curve decompositions}
The discrete QR basis $\Psi$ should not be confused with the principal component decomposition of historical yield data~\citep{litterman1991}. PCA orthogonalizes across time (diagonalizing the yield covariance matrix estimated from historical data), producing data-dependent factors typically interpreted as level, slope, and curvature. The QR-based orthogonalization presented here operates across maturities at a single date, producing model-based orthogonal directions that depend on the maturity grid $\tau$ and the decay parameters $\lambda$ but not on historical data. Both approaches produce uncorrelated ``modes,'' but they answer different questions: PCA identifies the dominant sources of yield curve variation over time, while QR orthogonalization decomposes the model-implied fit into independently estimable components at a given date.

\subsection{Computational implications for calibration workflows}

For repeated NSS calibration workflows, including central-bank term-structure production and fixed-income risk systems, the orthogonal-coordinate representation described in Section~\ref{sec:fit-gamma} applies directly: one computes $\hat{\gamma} = \Psi^\top\hat{y}$ from a QR decomposition and a matrix-vector product. The fitted yield curve $\hat{y} = \Psi\hat{\gamma}$ is identical to the classical result but is obtained without solving an ill-conditioned system. The economic parameters $\beta$ should only be recovered via $\hat{\beta} = R^{-1}\hat{\gamma}$ when $|R_{44}|$ is sufficiently large; otherwise, the individual values of $\hat{\beta}_3$ and $\hat{\beta}_4$ are numerically meaningless and only their sum $\hat{\beta}_3 + \hat{\beta}_4$ is reliable.

\paragraph{Separable theory and practical calibration}
Banholzer et~al.'s separation between outer nonlinear decay parameters~$\lambda$ and inner linear parameters~$\beta$ is mathematically correct for direct yield fitting. In many production environments, however, NSS is not calibrated as an explicit separable yield regression but as part of a broader nonlinear pricing stack, minimizing pricing errors across multiple instruments. Similar term-structured calibration problems also arise in option pricing; see Guterding and Boenkost~\citep{guterding2018} for calibration of a Heston model with piecewise constant time-dependent parameters to FX option market data. In such a setting the inner linear solve is often obscured or bypassed, and restartability becomes an operational requirement. The daily historical application in Section~\ref{sec:timeseries} indicates that the $\gamma$-based parametrization remains useful in this setting: warm-started daily recalibration produces stable orthogonal parameter series without changing the fitted curve.

\paragraph{Connection to dynamic Nelson--Siegel models}
In the dynamic Nelson--Siegel (DNS) framework~\citep{diebold2006,diebold2013}, the linear parameters $\beta_t$ are time-varying latent factors estimated via state-space methods (Kalman filter), while $\lambda$ is typically fixed. The measurement equation at each date $t$ is exactly the cross-sectional linear regression $\hat{y}_t = \Phi(\lambda)\beta_t + \varepsilon_t$ that the orthogonal reparametrization transforms. Applying the QR decomposition to $\Phi(\lambda)$ orthogonalizes the measurement update of the Kalman filter, potentially improving its numerical conditioning when $\lambda_1 \approx \lambda_2$. The $R_{44}$ diagnostic provides a time-invariant (for fixed $\lambda$) signal for whether the dynamic Svensson extension adds information beyond dynamic Nelson--Siegel. For the arbitrage-free DNS class~\citep{christensen2011}, the orthogonal parameters $\gamma_t$ inherit the no-arbitrage restrictions through $\gamma_t = R(\lambda)\beta_t$, preserving the economic structure. Recent extensions to time-varying decay parameters~\citep{caldeira2025} represent exactly the setting where the $R_{44}$ diagnostic would be most useful as a time-varying model selection criterion.

\subsection{Limitations and future work}

Most of our detailed conditioning and profile-likelihood analysis focuses on the linear subproblem (fixed $\lambda$), although Proposition~\ref{prop:fullcov} and Corollary~\ref{cor:beta-cov} extend the uncertainty discussion to the first-order jointly estimated case. The outer nonlinear optimization over $\lambda$ nevertheless remains non-convex. One extension would be to combine the orthogonal reparametrization with the penalty approach of~\citep{banholzer2024}. The real-data application in Section~\ref{sec:timeseries} is deliberately simple: it uses independent daily fits and an unweighted least-squares objective on a fixed reduced 9-tenor Treasury grid. Relevant extensions include weighted fits, alternative fixed tenor sets that may extend the usable history farther back, and targeted analyses of turmoil windows such as the oil shocks, the COVID episode, or the Ukraine-war period. Even in the exact changepoint setting, model-complexity selection remains open on this sample: the dynamic-programming segmentation is globally optimal conditional on the number of breaks, but a naive BIC still over-segments, so more refined penalties or Bayesian model-selection criteria would be worth investigating. The continuous analytical basis derived in Proposition~\ref{prop:finiteT} suggests another direction: comparing horizon-dependent diagnostics based on $L^2(0,T)$ with the exact grid-dependent QR diagnostic used in practice. Analyzing the gradient and Hessian of the reduced objective $H(\lambda)$ in the orthogonal basis---in particular, whether the orthogonal parametrization simplifies the Kaufman~\citep{kaufman1975} Jacobian approximation near the degenerate manifold---is another extension.

\section{Conclusion}\label{sec:conclusion}

We have analyzed orthogonal reparametrization of the NSS model as an exact computational reformulation of a structured separable nonlinear least-squares problem. Orthogonalization leaves the fitted curve unchanged, but it reveals the conditioning and identifiability structure of the inner problem: the orthogonal parameters $\gamma$ are uncorrelated with uniform conditional uncertainty, and when the decay parameters are estimated jointly their first-order full covariance takes an explicit Schur-complement form that isolates the additional nonlinear uncertainty. The QR factor $R_{44}$ serves as a scalar identifiability diagnostic, and the orthogonal basis provides a transparent framework for understanding the bias-variance trade-off in regularized estimation, though standard ridge regression remains preferable in the degenerate regime. Beyond the discrete QR basis used for fitting, we derived a grid-independent analytical orthogonalization on finite maturity horizons $[0,T]$ and showed that it yields a full $4\times4$ closed-form NSS basis without any artificial damping parameter. At the same time, the infinite-domain analysis clarifies a sharp limitation: on $(0,\infty)$ only the decaying $3\times3$ subspace admits a parameter-free orthogonalization, because the constant level mode does not belong to ordinary $L^2$. The synthetic experiments and the long-history Treasury calibration study show that this perspective is substantive rather than merely formal: on a fixed reduced maturity grid, the orthogonal parameter series $\gamma_t$ evolve more smoothly than the classical parameter series $\beta_t$ while the moving basis remains nearly constant, making $\gamma_t$ suitable as a historical state representation and as a cleaner lens on long-run regime shifts. A supplementary exact monthly changepoint analysis shows that this advantage is not an artifact of heuristic smoothing scores: $\beta$ and $\gamma$ imply almost the same broad regime dates, but $\gamma$ yields a materially cleaner segmentation of that common history. These results complement existing approaches based on penalization~\citep{banholzer2024} and parameter space restriction, and are readily implementable either as the QR-based inner linear solve in separable NSS fitters or as a post-fit diagnostic when an existing pipeline already supplies decay-parameter estimates.

\appendix
\numberwithin{equation}{section}
\numberwithin{figure}{section}
\numberwithin{table}{section}

\section{Analytical finite-horizon basis: closed forms}\label{app:finiteT-derivation}

For completeness, we collect here the closed-form expressions underlying Proposition~\ref{prop:finiteT}. Let $E_1(x) = \int_x^\infty e^{-u}u^{-1}\,du$ denote the exponential integral. For $a,b > 0$, define
\begin{align*}
I_T(a) &= \int_0^T e^{-a\tau}\,d\tau = \frac{1 - e^{-aT}}{a},\\
F_T(a) &= \int_0^T \frac{1 - e^{-a\tau}}{a\tau}\,d\tau
= \frac{\gamma_E + \ln(aT) + E_1(aT)}{a},\\
FE_T(a,b) &= \int_0^T \frac{1 - e^{-a\tau}}{a\tau}e^{-b\tau}\,d\tau
= \frac{\ln\!\bigl((a+b)/b\bigr) + E_1((a+b)T) - E_1(bT)}{a},
\end{align*}
and
\begin{equation*}
K_T(a,b) = \int_0^T
\frac{1 - e^{-a\tau}}{a\tau}
\frac{1 - e^{-b\tau}}{b\tau}\,d\tau
\end{equation*}
with closed form
\begin{equation*}
\begin{aligned}
K_T(a,b)
&= \frac{1}{ab}\Bigl[
a\ln \bigl((a+b)/a\bigr) + b\ln \bigl((a+b)/b\bigr) \\
&\qquad + (a+b)E_1((a+b)T) - aE_1(aT) - bE_1(bT) \\
&\qquad - \bigl(1 - e^{-aT} - e^{-bT} + e^{-(a+b)T}\bigr)/T
\Bigr].
\end{aligned}
\end{equation*}
Then, with $a=\lambda_1$ and $b=\lambda_2$, the entries of $G_T$ are
\begin{align*}
G_{11} &= T,\\
G_{12} &= F_T(a),\\
G_{13} &= F_T(a) - I_T(a),\\
G_{14} &= F_T(b) - I_T(b),\\
G_{22} &= K_T(a,a),\\
G_{23} &= K_T(a,a) - FE_T(a,a),\\
G_{24} &= K_T(a,b) - FE_T(a,b),\\
G_{33} &= K_T(a,a) - 2FE_T(a,a) + I_T(2a),\\
G_{34} &= K_T(a,b) - FE_T(a,b) - FE_T(b,a) + I_T(a+b),\\
G_{44} &= K_T(b,b) - 2FE_T(b,b) + I_T(2b),
\end{align*}
with $G_{ji}=G_{ij}$ by symmetry.

These formulas follow by expanding products of the NSS basis functions into combinations of $1$, $e^{-a\tau}$, $e^{-b\tau}$, and $\tau^{-1}(1-e^{-a\tau})$, then evaluating the resulting integrals term by term. The functions $F_T$, $FE_T$, and $K_T$ capture the only non-elementary pieces; their dependence on $E_1$ is exactly what makes the finite-horizon basis explicit but still closed form.

\section{Post-gap max-tenor robustness check}\label{app:tenor-robustness}

To isolate dependence on the number of maturities while holding the time window fixed after the 20Y outage, we reran the daily QR analysis on the common post-gap sample 1993-10-01 to 2026-04-02 using two fixed grids: the reduced 9-tenor grid from the main text and the maximal fixed 10-tenor grid obtained by reinstating 20Y. Both samples contain 8{,}480 daily curves. Table~\ref{tab:postgap-tenor-robustness} summarizes the comparison.

\begin{table}[tb]
\centering
\small
\caption{Common-window post-gap robustness check (1993-10-01 to 2026-04-02): reduced 9-tenor grid versus maximal fixed 10-tenor grid. For the component-wise smoothness diagnostics, entries are reported in component order $j=1,2,3,4$.}
\label{tab:postgap-tenor-robustness}
\begin{tabular}{lcc}
\toprule
Metric & Reduced 9-tenor & Maximal 10-tenor \\
\midrule
$\rho(\beta_j)/\rho(\gamma_j)$ & $2.39,\ 1.49,\ 2.96,\ 1.26$ & $1.93,\ 1.23,\ 0.66,\ 0.54$ \\
$J(\beta_j)/J(\gamma_j)$ & $2.15,\ 1.20,\ 2.92,\ 1.11$ & $1562,\ 724,\ 22.4,\ 198.5$ \\
Share with $|R_{44}|<0.1$ & $5.2\%$ & $100\%$ \\
$\min |R_{44}|$ & $2.62\times10^{-2}$ & $3.87\times10^{-6}$ \\
$q_{0.95}(B_t)$ & $5.02\times10^{-7}$ & $3.16\times10^{-5}$ \\
$\max B_t$ & $1.93$ & $1.53$ \\
Elbow-selected changes $(K_\beta, K_\gamma)$ & $(5,5)$ & $(1,6)$ \\
\bottomrule
\end{tabular}
\end{table}

Two observations are relevant. First, on this common post-gap window the reduced 9-tenor grid still reproduces the main qualitative conclusion: the orthogonal parameter series are smoother in components 1--3, the changepoint heuristic selects the same five-break complexity in both parametrizations, and at that common complexity the within-segment SSE is lower for $\gamma$ by about $22.5\%$.

Second, reinstating the 20Y tenor changes the problem qualitatively. The max-tenor grid is almost always in the weak-identification regime: $|R_{44}|<0.1$ on every day and can fall to $3.87\times10^{-6}$. On this grid the scale-normalized roughness ratio $\rho$ becomes hard to interpret because the classical parameter series themselves can blow up, so the jump statistic $J$ and the $R_{44}$ diagnostic are more informative. The changepoint comparison also ceases to be clean in the same sense as in the baseline sample: the elbow heuristic selects only one changepoint for $\beta$ but six for $\gamma$. Even the reduced 9-tenor sample is somewhat less benign here than in the main text because the warm-start chain is reinitialized at 1993-10-01 rather than inherited from the pre-gap history, but reinstating 20Y amplifies that instability dramatically.

These results support the reduced 9-tenor specification as the baseline long-history sample. The denser post-gap 10-tenor grid is still useful as a stress test, but it reintroduces the near-degenerate regime that the main specification was chosen to avoid.

\section*{Funding}

D.~G. acknowledges the support of the Investment Bank of the State of Brandenburg (ILB) under Grant No.~86001926, which was co-financed by the European Union through its regional development fund (ERDF).

\section*{Declaration of competing interest}

The authors declare that they have no known competing financial interests or personal relationships that could have appeared to influence the work reported in this paper.

\section*{Data availability}

The source code used to generate the numerical experiments, figures, and tables reported in this manuscript is publicly available on Zenodo (\url{https://doi.org/10.5281/zenodo.19676952}). The repository includes the scripts and pinned software dependencies required to reproduce the reported computational results, together with the generated intermediate data products needed for the reported figures and tables. The long-history U.S.\ Treasury calibration study uses publicly available yield-curve data documented by G\"urkaynak et~al.~\citep{gurkaynak2007} and downloaded from the U.S.\ Federal Reserve's H.15 statistical release at \url{https://www.federalreserve.gov/releases/h15/}; synthetic experiments are generated directly from the included code. A versioned archival snapshot suitable for citation will be added before submission of the final accepted version.

\section*{Acknowledgments}

The research questions, mathematical framework, interpretation of results, and all scientific conclusions are the authors' own contribution. AI assistants (Claude, Opus~4.6, Anthropic, and GPT-5.4, OpenAI, via Codex) were used during the research process for literature surveying, code drafting, figure generation scripting, and manuscript editing. All mathematical derivations, numerical results, and claims were independently verified by the authors. The iterative workflow follows agent-driven scientific-computing methodologies~\citep{resnik2025}.

\end{document}